\documentclass[12pt]{iopart}
\usepackage{iopams}
\usepackage{graphicx}
\pdfoutput=1

\def\be{\begin{equation}}
\def\ee{\end{equation}}
\def\ba{\begin{eqnarray}}
\def\ea{\end{eqnarray}}

\def\ov{\overline}

\def\Z{\mathbb{Z}}

\def\nl{\nonumber \\}

\def\ap{\approx}

\def\ra{\rangle}

\def\la{\langle}
\def\de{\partial}

\def\wh{\widehat}
\def\Tr{{\rm Tr}}
\def\dag{\dagger}

\def\a{\alpha}

\def\g{\gamma}
\def\G{\Gamma}
\def\D{\Delta}
\def\d{\delta}
\def\e{\epsilon}
\def\eps{\varepsilon}

\def\th{\theta}

\def\l{\lambda}
\def\L{\Lambda}

\def\p{\pi}
\def\P{\Pi}
\def\r{\rho}
\def\s{\sigma}

\def\f{\varphi}

\def\w{\omega}

\def\B{{\bf B}}

\begin{document}

\title[Matrix Effective Theories of the Fractional Quantum Hall effect]
{Matrix Effective Theories of the Fractional Quantum Hall effect}

\author{Andrea Cappelli and Ivan D. Rodriguez}

\address{I.N.F.N. and Dipartimento di Fisica\\
Via G. Sansone 1, 50019 Sesto Fiorentino - Firenze, Italy}
\ead{andrea.cappelli@fi.infn.it, rodriguez@fi.infn.it}

\begin{abstract}
The present understanding of
nonperturbative ground states in the fractional quantum Hall effect
is based on effective theories of the Jain ``composite fermion'' excitations.
We review the approach based on matrix variables, i.e. D0 branes,
originally introduced by Susskind and Polychronakos.
We show that the Maxwell-Chern-Simons matrix gauge theory
provides a matrix generalization of the quantum Hall effect,
where the composite-fermion construction naturally follows from
gauge invariance. The matrix ground states obtained
by suitable projections of higher Landau levels are found to
be in one-to-one correspondence with
the Laughlin and Jain hierarchical states.
The matrix theory possesses a physical limit for commuting matrices
that could be reachable while staying in the same phase.
\end{abstract}

\maketitle

%-1--------------------------------------------
\section{Introduction}

\begin{figure}
\includegraphics[width=14cm]{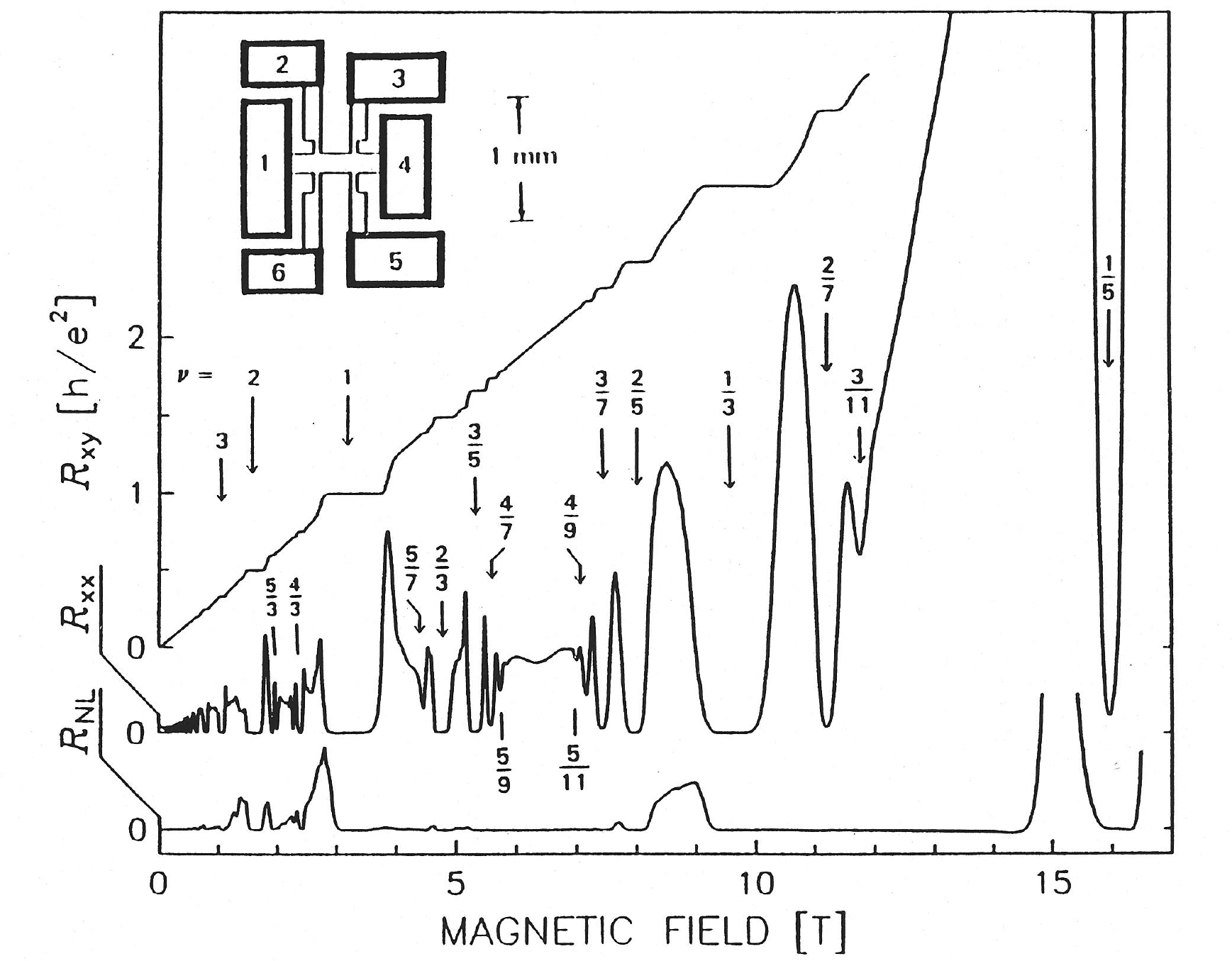}
\caption{Diagonal and transverse resistivities in the fractional 
quantum Hall effect \cite{jain}.}
\label{sampleFQHE}
\end{figure}

The quantized Hall effect occurs in systems of planar electrons
inside layered semiconductors, that are placed
in strong magnetic fields $\textbf{B}$ ($\sim 10$ Tesla)
and very low temperatures ($T\sim 1 {\rm mK} - 1 {\rm K}$)
\cite{prange}. 
For certain values of the field, the longitudinal Ohmic current 
vanishes and the transverse component $R_{xy}$ of the resistivity 
(Hall resistivity) is quantized (Fig.\ref{sampleFQHE}):
\ba
 R_{xy} &=& \sigma_{xy}^{-1}=\nu^{-1} \frac{h}{e^2} , 
\quad \nu=1,2,3,...,\frac{1}{3},\frac{1}{5},...,\frac{2}{5},
\frac{2}{7},\frac{3}{5},..., \nl 
R_{xx} &=& \sigma_{xx}= 0 , \
\label{resist}
\ea
where $\nu$ is the ``filling fraction'', that can be integer or fractional  
\cite{tsui}. 
The regimes in which the values of the resistivity are given by 
(\ref{resist}) are called ``plateaux'' of the Quantum
Hall effect (QHE). They correspond to very stable gapful ground states 
with uniform density 
$\overline{\rho} =\nu e\textbf{B}/hc$, where the
electrons behaves like a fluid with characteristic quantum effects 
\cite{prange}. 
The low-energy excitations are local deformations in the 
density (vortices) called quasi-holes and quasi-particles; 
the density waves are also gapful such that the quantum fluid is incompressible.
The integer Hall effect can be described in terms of free electrons filling up 
the Landau levels, while the fractional effect
requires to consider interacting electrons.

In 1983, Laughlin proposed a phenomenological theory for the
fillings $\nu=1/(2k+1)$, with $k$  positive integer  \cite{laugh}: 
he described the incompressible fluid and predicted quasi-particles 
with fractional charge that were observed in 1997 \cite{lau-experim}.
Other filling fractions not described by Laughlin's theory are observed
experimentally, belonging to the series
$\nu=n/(2nk\pm 1)$, where $n>1$ and $k$ are positive integers \cite{prange}. 
Upon introducing the idea of ``composite fermions'' excitations,
Jain argued that these fractional quantum Hall states actually correspond to
integer quantum Hall states of composite fermions \cite{jain}. 
Based on this relation, Jain obtained trial
wavefunctions that are confirmed by the numerical analyses. 
Moreover, weakly-interacting composite-fermion excitations 
have been observed in several experiments \cite{prange}. 
Fradkin and Lopez \cite{fradkin} and others \cite{cftheories} 
realized the Jain correspondence in
quantum field theory by letting the electrons to interact 
with a ``statistical'' Chern-Simons gauge field. 
They studied the theory within the mean field approximation and 
reproduced the Jain ground states and some of their phenomenological 
features.

%-f1-------------------------------------

In this contribution, we review another possible effective theory for 
the fractional QHE, that is based on matrix models or, more precisely, on  
gauge theories of matrices in $(2+1)$ dimensions, that are
equivalent to noncommutative gauge theories.
This approach is not yet fully developed, but it presents some interesting
features that we believe are worth discussing.

The presentation is organized as follows:
the next section contains a short introduction to 
the phenomenology of the integer and fractional QHE. 
We review the Laughlin theory \cite{laugh}, the Jain interpretation of 
the fractional QHE \cite{jain} and the field theory proposed by 
Fradkin and Lopez \cite{fradkin}.
The third section deals with the Chern-Simons matrix model, and reviews 
the work by Susskind and Polychronakos.
Using results for D0-branes in string theory, Susskind showed
that two-dimensional semiclassical incompressible fluids in strong
magnetic field could be described by the noncommutative Chern-Simons theory
\cite{susskind}. 
Indeed, the use of noncommuting spatial coordinates, 
$x_1$, $x_2$, i.e. $\left[ x_1, x_2 \right] = i\theta$, 
implies a generalized uncertainty relation that controls the effective size 
of electrons and thus reduces the density of the fluid, 
leading to $\nu=1/(1+{\bf B}\theta) < 1$.
Afterwards, Polychronakos modified the theory to describe a finite droplet 
of fluid, and obtained the $U(N)$ matrix gauge theory called Chern-Simons 
matrix model \cite{poly1}. 
From the quantization of this theory, one finds the important result 
that the ground states are exactly given by the Laughlin wave functions 
\cite{heller}\cite{cr}. However, the Chern-Simons matrix model does not 
naturally describe the more general Jain states
and the full quantum theory does not reproduce the electron system 
of the QHE \cite{cr1}\cite{hansson2}.

Section 4 is devoted to our proposal of the Maxwell-Chern-Simons matrix 
theory \cite{cr1}: this is a generalization of Susskind-Polychronakos theory 
that contains an additional coupling $g\ge 0$ controlling matrix 
noncommutativity. 
At $g=0$, the theory corresponds to a matrix generalization 
of the Landau levels, with an exponentially growing density of states
that is typical of matrix theories.
We introduce a set of projections that not only limit the degeneracy 
but also uniquely selects ground states that
are matrix analogs of the expected Laughlin and Jain states (section 5). 
This is the most interesting feature of the matrix approach to the QHE,
namely that the phenomenological ground states arise naturally 
from gauge invariance and the projections.
The same ground states are also found in the 
semiclassical analysis of the theory \cite{cr2}: they
correspond to incompressible fluids with piecewise constant density, 
as expected \cite{jain}. 
In section 6, we discuss the Maxwell-Chern-Simons matrix theory for $g>0$:
in the $g = \infty$ limit, the matrix coordinates commute and
the theory describes ordinary electrons in Landau 
levels with $O(1/r^2)$ interaction, which is a good
approximation of the QHE system \cite{prange}.
Although the phase diagram ($0<g<\infty$) of the Maxwell-Chern-Simons 
theory is not yet known, we conjecture
that the matrix ground states found at $g=0$ have a smooth $g\to \infty$ 
limit into the phenomenological Laughlin and Jain states 
(no phase transition for finite $g$ values) \cite{cr1}. 
The proof of this fact would confirm the physical relevance 
of the matrix theory approach to the fractional QHE. 
In the conclusions (section 7), we discuss some developments 
of this line of research.

This paper is dedicated to the vivid memory of Alyosha Zamolodchikov.

%-2--------------------------------------------------
\section{Review of the Fractional Quantum Hall Effect}
\subsection{Landau levels}

Consider planar electrons of mass $m$ and electric charge $e$ in an
uniform magnetic field $\textbf{B}$ (in units $\hbar= 1 $,
$c = 1$). The one-particle Hamiltonian is given by:
\ba 
H = -\frac{1}{2m} (\nabla - ie\textbf{A})^2 . \ 
\label{landau-hamilt} 
\ea
We work in the symmetric gauge for the vector potential,
$A_i = (\textbf{B}/2) \epsilon_{ij}x^j, \ i,j= 1,2$. 
The magnetic field introduces a length scale, the so-called magnetic
length, $\ell = \sqrt{2/e\textbf{B}}$.
The use of holomorphic spatial coordinates $z = x_1 + ix_2$ and 
$\overline{z} = x_1 - ix_2$,  is natural in the QHE \cite{winf}. 
By introducing two commuting sets of harmonic oscillators 
($\partial = \frac{\partial}{\partial z}$ and 
$\overline{\partial} = \frac{\partial}{\partial\overline{z}}$),
\ba
d&=&\frac{z}{2 \ell} + \ell \overline{\partial} \ , 
\qquad d^\dag=\frac{\overline{z}}{2 \ell} - \ell
\partial \ , \qquad \left[ d , d^\dag \right]=1 , \nl 
c&=&\frac{\overline{z}}{2 \ell} + \ell
\partial \ , \qquad c^\dag = \frac{z}{2 \ell} - \ell \overline{\partial}
\ , \qquad \left[ c , c^\dag \right] =1 , \ 
\label{landau-comm}
\ea
the Hamiltonian (\ref{landau-hamilt}) and
the angular momentum can be written as follows:
\ba 
H &=& \omega \ ( d^\dag d + \frac{1}{2}) , \nl 
J &=& c^\dag c - d^\dag d , \label{energ-landau} 
\ea
where $\omega = e\textbf{B}/m$ is the cyclotron frequency. 
Since the operators $c$ and $d$ commute, the spectrum consists of 
infinitely degenerate levels ($c^\dag c$ excitations) with energies
$\epsilon_n =\omega n$, i.e. the Landau levels ($d^\dag d$ ladder). 
The degenerate states correspond to the semiclassical cyclotron orbits,
that are quantized by the condition that the contained flux 
is a multiple of the quantum unit $ \phi_0=eh/c$, i.e. $BA_j=j \phi_0$,
with $j$ the angular momentum eigenvalue.
In the lowest Landau level, the one-particle wave functions take the form:
\ba
\f_j(z,\ov{z})= \frac{1}{\ell \sqrt{\pi}} 
\frac{1}{j!} \left( \frac{z}{\ell} \right)^{j} e^{-z\ov{z} / (2\ell^2) } \ , 
\qquad d\ \f_j(z,\ov{z})=0,
\label{low-LL}
\ea
i.e. they are holomorphic in $z$ up to an exponential factor.
The associated one-particle densities are indeed peaked at the
semiclassical orbits.

\begin{figure}
\begin{center}
\includegraphics[width=10cm]{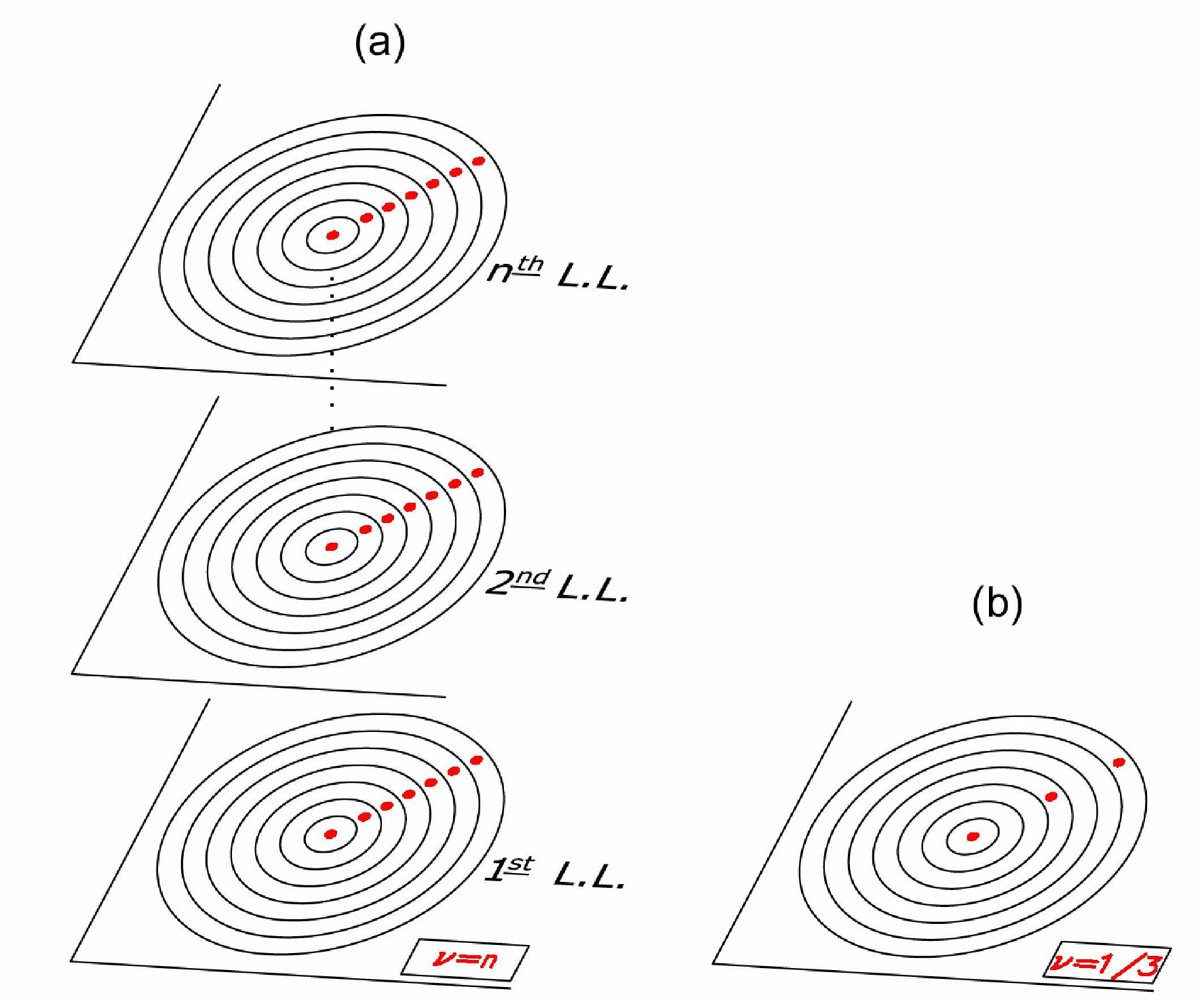}
\caption{Graphical representation of the Landau levels: (a) integer 
and (b) fractional filling.}
\label{LandLev}
\end{center}
\end{figure}

On a finite region of area $A$, the number of degenerate states is equal 
to the flux through the system in quantum units, $N_\phi={\bf B} A/\phi_0$.
In completely filled Landau levels, the Hall conductivity is given by 
$\s_{xy}=R_{xy}^{-1}=\nu \ e^2/h$, where $\nu=N/N_\phi$ is 
the filling fraction, i.e. the number of electrons $N$ divided 
by the number of available states. Figure \ref{LandLev}(a) 
shows the $\nu=n$ case in which $n$ levels are filled with one electron
 per orbital (the spin degree of freedom is frozen in the direction of 
$\bf{B}$). The density is uniform and the electron fluid is incompressible 
due to the exclusion principle, the gap being given by $\omega$. 
Thus the simple theory of free electrons in Landau levels is sufficient 
to describe the main physical properties of the integer QHE.
(The formation of the plateaux near integer fillings is
explained by the localization of excitations due to disorder) \cite{prange}.

On the other hand, if there are many empty orbitals like in the case $\nu=1/3$
(Fig.\ref{LandLev}(b)), the free-electron 
states are compressible, in contrast with the experimental observation:
the fractional QHE requires the study of interacting electrons.
The formation of the gap by the Coulomb potential is clearly nonperturbative:
one should try an approach based on ansatzs and effective theories,
supplemented by numerical analyses.
It turns out that the ground states are condensates of 
charges and vortices which have some analogies with superfluids
and confined gauge theories, but are also specific of the
two-dimensional parity breaking setting.

%-f2----------------------------------

%-2.2--------------------------------------------------
\subsection{The Laughlin theory}

In a remarkable paper \cite{laugh}, Laughlin proposed a class of 
trial wave functions given by:
\ba
\Psi_{m}(z_1,z_2,...,z_N)=
\prod_{i<j=1}^N (z_i-z_j)^m e^{-\frac{1}{2} \sum_i^N \mid z_i \mid^2 } \ ,
\label{laugh-wfunc} \ea
with $N$ the number of electrons and $m$ an integer parameter. 
Hereafter we set the magnetic length $\ell=1$. 
The wave functions (\ref{laugh-wfunc}) describe spinless electrons 
in the lowest Landau level: $m$ must be odd,
$m=2k+1$, for antisymmetry of fermions.
In order to determine the properties of these states, 
Laughlin used the analogy with a two-dimensional plasma, as it follows.
The determination of the one-particle density can be reduced 
to the analysis of the two-dimensional statistical model of charges defined by,
\ba
Z_{plasma}&=&
\parallel \Psi_{m} \parallel^2 = 
\int \prod_{i=1}^N d^2z_i \ e^{-\beta H_{plasma}} \ , \nl 
H_{plasma}&=& m \sum_{i=1}^N \mid z_i \mid^2 - m^2 \sum_{i<j=1}^N 
log \mid z_i - z_j \mid^2 \ .
\label{plasma-analogy}
\ea
In this equation, $H_{plasma}$ describes a classical plasma of charges 
in a uniform background at temperature $\beta=1/m$. 
Knowing that this plasma is totally screened for small values of $m$, 
Laughlin could argue that the density is uniform and could calculate 
the gap of excitations. For constant density, the parameter $m$ 
can be readily related to the filling fraction by $\nu=1/m$.
Note that Laughlin's wavefunctions vanishes as $\left( z_i - z_j \right) ^m$ 
when any two particles $i$ and $j$ approach each other: namely,
the amplitude for nearby particles is very small and the expectation value 
of the Coulomb energy is consequently reduced. This is a rather successful
property from the variational point of view, since other wave functions 
with same average density do have this feature. 
Numerical experiments show that the Laughlin wavefunction is actually 
very close to the exact ground state for several short-range repulsive 
interactions \cite{prange}\cite{laugh}\cite{numeric}.

\begin{figure}
\begin{center}
\includegraphics[width=9cm]{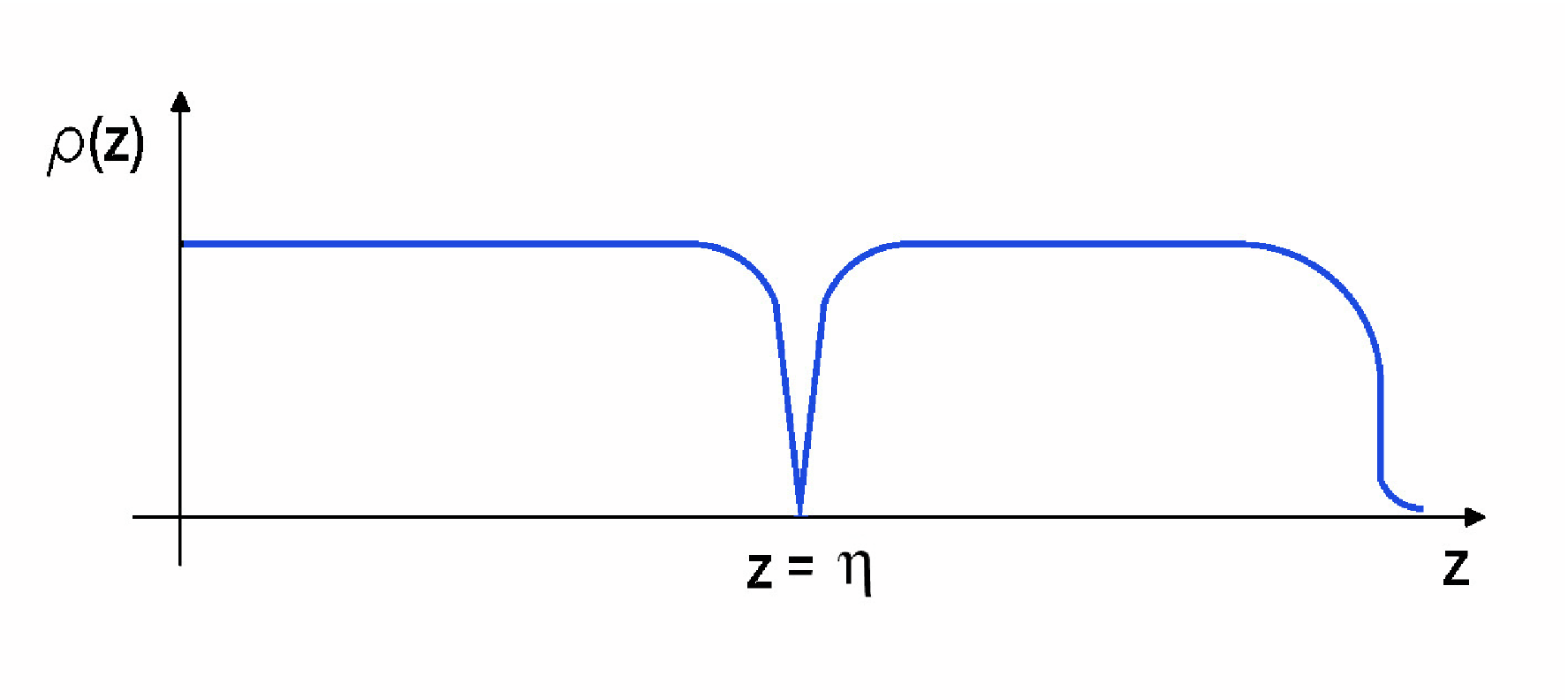}
\caption{Schematic plot of the density of the electron fluid
in presence of a quasi-hole at $z=\eta$.}
\label{plotqh}
\end{center}
\end{figure}

Laughlin also proposed the wave functions of the low-energy 
quasiholes excitations: they are localized density deformations,
\be
\Psi_{qh}=(\eta; z_1,...,z_N) = \prod_{i=1}^N (\eta - z_i) 
\prod_{i<j=1}^N (z_i - z_j)^{2k+1} e^{-\frac{1}{2
\ell^2}\sum_i \mid z_i \mid^2}, \ 
\label{q-hole} 
\ee
with $\eta$ the position of the vortex (Fig.\ref{plotqh}). 
To calculate the charge of the quasi-hole, one can use 
the plasma analogy (\ref{plasma-analogy}):
\be
\parallel \Psi_{qh} \parallel^2=\int \prod_{i=1}^N d^2 z_i 
e^{- \beta \left( m \sum_i \mid z_i \mid^2 - 
m^2 \sum_{i<j} log \mid z_i - z_j \mid^2 - m \sum_i log \mid z_i - \eta \mid
\right)}. 
\label{charge-qh}
\ee
Comparing (\ref{charge-qh}) with (\ref{plasma-analogy}), 
we observe that the electrons feel the presence of a
charge $1/m$ at the point $z=\eta$: 
thus, the quasi-holes have fractional charge $Q_{qh}=e/m$ \cite{laugh}.

%-f3-------------------------------------

In the wave function for two quasi-holes,
\be
\Psi_{2 qh}(\eta_1,\eta_2; z_1,...,z_N) = (\eta_1 - \eta_2) ^{\frac{1}{2k+1}}
 \prod_i (\eta_1 - z_i) \prod_i
(\eta_2 - z_i) \Psi_{m},\ 
\label{quasihole} 
\ee
Laughlin introduced the term $(\eta_1 - \eta_2)$ raised to a fractional power,
that is necessary for charge equilibration in the plasma 
(\ref{charge-qh}); he assumed a holomorphic dependence
as for the electron coordinates \cite{wilczek}.
If we rotate one quasi-hole around the other, we obtain:
\be
\Psi_{2 qh}((\eta_1 - \eta_2)\rightarrow
e^{i\pi}(\eta_1-\eta_2))= e^{i\frac{\pi}{2k+1}}\ \Psi_{2 qh}(\eta_1,\eta_2).
\label{statist-qh}
\ee
Therefore, the wave function acquires a non-trivial phase under 
exchanges of excitations: the quasi-holes have 
``fractional statistics'',  $\frac{\theta}{\pi}=\frac{1}{2k+1}$.
This is a long-range, topological interaction of vortices, that is
allowed in parity breaking two-dimensional systems \cite{wilczek}.
The fractional charge and statistics of excitations are confirmed 
by the effective field theory descriptions to be discussed later. 
The fractional charge has been observed in experiments of 
quasiparticle tunneling \cite{lau-experim};
the fractional statistic has not been detected directly but 
there are indirect confirmations \cite{prange}.

%-2.3-----------------------------------

\subsection{The Jain interpretation of the fractional 
quantum Hall effect}

In Fig. \ref{sampleFQHE}, one finds stable plateaux 
at other filling fractions that are nicely accounted by 
the series $\nu=n/(2nk\pm 1)$ with $n>1$. 
The phenomenological theory due to Jain explains them
as follows: the argument starts by observing that the inverse filling,
\be
\frac{1}{\nu} = \frac{N_\phi}{N} =\pm\frac{1}{n} + 2 k\ ,
\label{jain-shift}
\ee
is equal the number of fluxes per electron. 
Imagine that it is possible to bound an even number of fluxes, i.e. $2k$, 
to each electron, to form a new quasiparticle called ``composite fermion''.
(Note that an even number of flux quanta yield an integer Aharonov-Bohm phase
and keeps the fermionic statistics). 
When $2k$  fluxes are attached to
each electron, the same number of fluxes are removed from 
the external magnetic field: therefore, the filling fraction of
the system of composite fermions is given by
\be
\!\!\!\!\frac{1}{\nu^*}=\frac{N_\Phi - 2k N}{N} =\frac{1}{\nu} - 2k= 
\pm\frac{1}{n}\ , 
\qquad
\textbf{B}^\star =\textbf{B} - 2k\ 2\pi\rho_0 \ ,
\label{nu-star}
\ee
corresponding to an integer QHE.  The reduced magnetic field felt by the 
new particles is given by $\textbf{B}^\star$.
This is indeed observed: many experiments confirm the 
existence of weakly interacting excitations feeling the reduced 
magnetic field, i.e. behaving as Jain's composite fermions \cite{prange}. 
The incompressibility of the fractional QHE is explained by Jain as due to 
the equivalence between the system of electrons
with $\nu= n/(2nk+1)$ and the integer QHE of composite fermions with 
$\nu^\star= n$.
 
Following the Jain argument, the flux attachment is clear in the form
of the Laughlin wave function ($\nu^\star =1$):  
the factor $\Pi ^N _ {i<j} (z_i - z_j)^ {2k}$ yields Aharonov-Bohm
phase of $2k$ flux quanta to any electron, and the rest is 
the wave function of the filled first Landau level. 
In the general case of $\nu= n/(2nk+1)$, the ground state wave functions 
proposed by Jain on the basis of his equivalence are:
\ba
\Psi_{ 2k+ \frac{1}{n} } (  z_1,...,z_N ) = 
\prod ^N_ {i<j} (z_i - z_j) ^ {2k} \Psi_{\frac{1}{n}} (z_1,...,z_N ), \
\label{jain-elecstate}
\ea
with $\Psi_{1/n}( z_1,...,z_N )$ being the wave functions with $n$ 
completely filled levels (Slater determinants). 
The Jain wave functions (\ref{jain-elecstate}) have been confirmed 
by comparison with numerical results of exact diagonalization 
of the Hamiltonian with Coulomb interaction 
\cite{jain}. The fillings $\nu= n/(2nk-1)$ are also described
by (\ref{jain-elecstate}) with charge-conjugate term
$\Psi_{1/n}\to \ov{\Psi}_{1/n}$. 

The Jain scheme also provide excellent approximations for 
the quasi-holes and quasi-particles excitations \cite{jain}.
For instance, a quasi-particle in the origin for the Jain
state with $\nu= n/(2nk+1)$ is given by,
\be
\Psi_{qp;\ 2k+ \frac{1}{n} } ( z_1,...,z_N
) = \prod ^N_ {i<j} (z_i - z_j) ^ {2k} 
\Psi_{qp; \frac{1}{n}}( z_1,...,z_N ), \ 
\label{q-hole-jain}
\ee
where $\Psi_{qp; 1/n}(  z_1,...,z_N )$  corresponds to the wave function 
of $n$ filled Landau levels and one electron in the first orbital 
of the $(n+1)$-th Landau level. 
The corresponding localized density has an excess of charge at the 
origin of the droplet.

%-2.4-----------------------------------------------------
\subsection{Fermion Chern-Simons field theory}

Among the effective field theories that have been proposed to 
describe the fractional QHE, we recall the theory of non-relativistic 
fermions coupled to the Abelian Chern-Simons ``statistical'' interaction, 
that has been developed by Fradkin and Lopez \cite{fradkin} 
and others \cite{cftheories}. The action can be schematically written:
\be
S=\frac{\kappa}{4\pi}\int\eps^{\mu\nu\rho} a_\mu \partial_\nu a_\rho + 
\int J_\mu a_\mu + S_{\mbox{fermion}} \ .
\ee
 Consider the Gauss law of this theory:
\be
j_0(\vec{x})=-\frac{\kappa}{2\pi} \mathcal{B}(\vec{x})=
-\frac{\kappa}{2\pi} \epsilon^{ij} \partial_i a_j(\vec{x}) \ ,
\label{c-s-gauss}
\ee
where $\mathcal{B}$ is the ``statistical'' magnetic field and 
$\kappa$ the Chern-Simons coupling constant.
At quantum level, this is an operator constraint which selects 
the physical space of states. These are charge-flux composites: 
every particle with unit electric charge carries a magnetic charge equal 
to $2\pi/\kappa$. The wave functions for these composite particles 
exhibits Aharonov-Bohm effects changing the statistics. 
Therefore a fermion coupled to a Chern-Simons gauge field behaves 
like an anyon with statistical angle $\theta/\pi = 1/\kappa$, 
measured with respect to the Fermi statistics \cite{fradkin}. 
If $\kappa = 1/2k$, where $k$ is an integer, then $\theta/\pi = 2 k$ 
and the composite states are still fermions.

A ground state with uniform density 
$\langle j_0(z) \rangle=\bar{\rho}$ implies a constant field:
\be
\langle \mathcal{B} \rangle = - \frac{2\pi\ov{\rho}}{\kappa}=
-2k\ 2\pi \rho_0 \ ,  
\qquad\quad
\langle \mathcal{E} \rangle = 0 \ . 
\label{Fradfluid}
\ee
Eq. (\ref{Fradfluid}) shows that, within the mean field approximation, 
the effect of the statistical
gauge field is to change the effective flux experienced by the
fermionic excitations. The effective magnetic field is 
$B_{eff}= \textbf{B} + \langle \mathcal{B} \rangle = 
\textbf{B} - 2 \kappa 2\pi  \rho_0$, 
in agreement with Jain's argument (\ref{nu-star}).

The uniform effective magnetic field $B_{eff}$ 
defines a new set of effective Landau levels. 
Each level has a degeneracy equal to the total number of effective flux quanta
$N_{eff}$ and the separation between levels is the effective 
cyclotron frequency $\omega_{eff}=|B_{eff}|/m$. 
Similarly, there is an effective cyclotron radius $\ell_{eff}$. 
It is easy to see that the mean-field approximation (uniform density)
(\ref{Fradfluid}) is selfconsistent only if the fermions fill exactly 
an integer number $n$ of effective Landau levels. 
This reproduces Jain's theory: the fractional QHE is the integer QHE of 
a system of electrons dressed by an even number of flux quanta. 
The allowed filling fractions are those obtained by Jain: 
$\nu=n/(2nk \pm 1)$ \cite{fradkin}.
Further results of this approach are reviewed in \cite{fradkin2}.

Let us mention for completeness the effective field theory approaches
based on $(1+1)$ dimensional conformal field theories.
As originally observed by Wen \cite{wen}, a droplet of incompressible fluid
possess low-energy massless chiral excitations at the edge, 
that can be described by conformal field theories with $U(1)$ current, 
also called chiral Luttinger liquid theories, and their generalizations.
These theories of the QHE have been extensively developed in the last
20 years and can describe the low-energy physics occurring in conduction 
experiments \cite{prange} \cite{wen} \cite{jainedge}. 
In this approach, the formation of the incompressible fluid
is assumed and cannot be derived, since the dimensional reduction is
only possible for these specific states; 
actually, there is a different conformal 
field theory for each plateaux, whose form can be inferred by the 
properties of excitations and other data. 
In the following, we deal with $(2+1)$-dimensional effective theories 
that could explain the formation of incompressible ground states.

%-3------------------------------------------
\section{Semiclassical incompressible fluid and noncommutative
Chern-Simons theory}

In this section we introduce the effective theories of the fractional QHE based
on matrix degrees of freedom, equivalent to noncommutative field theories.
The subject was initiated by Susskind in 2001, who observed the analogies
between the QHE and the physics of D-branes in string theory
\cite{susskind}\cite{dzero}.
We shall find that the matrix d.o.f. have associated a gauge field
and their Gauss law provides another realization of Jain's flux attachment
to particles (\ref{nu-star}). 
The description of the QHE by matrix theories is far less developed than 
the fermion Chern-Simons theory, but there are some nice features like the
explicit relation with wave functions; in our opinion, the matrix theories
could provide another view on the physics of composite fermions.

We start by reviewing Susskind's work \cite{susskind}, 
who observed that the semiclassical limit of 
noncommutative Chern-Simons field theory
could describe incompressible fluids in high magnetic 
field with Laughlin's filling fractions \cite{susskind}.
Consider $N$ first-quantized electrons with two-dimensional 
coordinates $X^a_\a(t)$, $a=1,2$, $\a=1,\dots,N$, placed in a strong
magnetic field $\bf{B}$, such that their action can be projected to 
the lowest Landau level \cite{dunne},
\be
L\ =\ \frac{e\bf{B}}{2} \sum_{\a=1}^N\ \e_{ab}\ X^a_\a\ \dot{X}^b_\a
\ .
\label{S-LLL}
\ee
We now consider the limit of the particle forming a continuous fluid:
\be
\vec{X}_\a(t)\ \to\ \vec{X}(\vec{x},t)\ ,
\qquad\qquad \vec{X}(\vec{x},t=0)\ =\ \vec{x}\ ,
\label{lag-fluid}
\ee
where $\vec{x}$ are the coordinates of an initial, reference 
configuration of the fluid.
The resulting fluid mechanics is in the Lagrangian formulation,
because the field $\vec{X}(\vec{x})$ follows the motion of the fluid 
\cite{jackiw}.
For incompressible fluids, the constraint of constant density,
$\r(\vec{X})=\r_o$, can be written in terms of Poisson brackets
$\{\cdot,\cdot\}$ of the $\vec{x}$ coordinate as follows:
\be
\r_o\ =\ \r(\vec{x})\ =
\r_o\left\vert\frac{\de\vec{X}}{\de\vec{x}}\right\vert \ =\
\frac{\r_o}{2}\ \e_{ab}\ \{X^a,X^b\}\ .
\label{incompr}
\ee
This constraint can be added to the Lagrangian by using
the Lagrange multiplier $A_0$,
\be
L\ =\ \frac{e\bf{B}\r_o}{2} \int d^2 x
\left[\e_{ab}\ X^a\left(\dot X^b -\th\{X^b,A_0\}\right)
\ + \ 2\th\ A_0 \right]\ ;
\label{S-winf}
\ee
in this equation, we introduced the constant $\th$,
\be
\th \ =\ \frac{1}{2\pi\r_o}\ ,
\label{theta-def}
\ee
that will later parameterize the non-commutativity.

The Lagrangian (\ref{S-winf}) is left invariant by reparametrizations
of the $\vec{x}$
variable with unit Jacobian, the area-preserving diffeomorphism,
also called $w_\infty$ transformations \cite{winf}: they
correspond to changes of the original labels of the fluid at $t=0$
(cf. Eq.(\ref{lag-fluid})) \cite{susskind}\cite{jackiw}.
The $w_\infty$ symmetry can be put into the form of a gauge invariance
by introducing the gauge potential $\vec{A}$, as follows:
\be
X^a\ =\ x^a +\ \th\ \e_{ab}\ A_b(x)\ ,
\label{gauge-def}
\ee
and by rewriting the Lagrangian (\ref{S-winf}) in the Chern-Simons form:
\be
L\ =\ -\frac{k}{4\pi} \int d^2x\ \e_{\mu\nu\r}\left(
\de_\mu A_\nu A_\r \ +\ \frac{\th}{3}\left\{A_\mu,A_\nu\right\}A_\r
\right)\ ,
\label{w-CS}
\ee
where  $A_\mu=(A_0,A_a)$ is the three-dimensional gauge field. 
The coupling constant $k$ parameterizes the filling fraction of the
semiclassical fluid:
\be
\nu^{(cl)}\ = \ \frac{2\p\r_o}{e\bf{B}}\ = \ \frac{1}{e\bf{B}\th} =
\ \frac{1}{k}.
\label{nu-class}
\ee

Based on the result (\ref{w-CS}), Susskind conjectured that the
non-commutative (Abelian) Chern-Simons theory could be
the complete theory going beyond the continuous
fluid approximation, i.e. accounting for the granularity of the electrons.
Its action is \cite{NCFT}:
\be
L_{NCCS}\ =\ -\frac{k}{4\pi} \int d^2x\ \e_{\mu\nu\r}\left(
\de_\mu A_\nu \star A_\r \ -\ \frac{2i}{3}A_\mu\star A_\nu\star A_\r
\right)\ ,
\label{noncommchern}
\ee
involving the Moyal star product:
\ba
(F \star G)(x_1,x_2)=F(x_1,x_2) 
e^{\frac{i\theta}{2}(\overleftarrow{\partial_{x_1}} 
\overrightarrow{\partial_{x_2}} -
\overleftarrow{\partial_{x_2}}\overrightarrow{\partial_{x_1}})} G(x_1,x_2) \ .
\label{moyal}
\ea
Actually, the two Lagrangians (\ref{noncommchern}) and (\ref{w-CS})
agree to leading order in $\theta$, i.e. for dense fluids.
In the new Lagrangian (\ref{noncommchern}), the gauge fields with Moyal product
have become Wigner functions (see next section) of the
noncommuting operators, $\wh{x}_1,\wh{x}_2$, the former spatial
coordinates:
\be
\left[\wh{x}_1,\wh{x}_2 \right] = x_1\star x_2 -x_2\star x_1=i\ \th \ .
\label{x-comm}
\ee
The corresponding quantization of the area can be thought of as a
discretization of the fluid (at the classical level),
with the minimal area $\th$ allocated to a single electron 
(cf.(\ref{nu-class})).

%-3.1------------------------------------------------
\subsection{Matrix representation of the 
noncommutative Chern-Simons theory}

Every noncommutative theory is equivalent to a matrix theory, 
with matrices of infinite order ($N\to\infty$) \cite{NCFT};
in particular, the noncommutative Chern-Simons theory (\ref{noncommchern}) 
is equivalent to the matrix theory \cite{seiberg}:
\ba
L = \frac{B}{2}
\Tr\left[ \ \eps_{ij} \ X_i(t)\ D_t \ X_j(t) +\ 2\th\ A_0(t) \right] ,
\label{mcs-action1}
\ea
where now $X_1(t),X_2(t)$ and $A_0(t)$ are $N \times N$ matrices 
($N \rightarrow \infty$) and the covariant derivative is 
$D_t X_j = \dot{X_j} - i \left[ A_0 , X_j \right]$ , $j=1,2$. 

The proof of the correspondence is simpler if we go from 
the matrix (\ref{mcs-action1})  
to the noncommutative theory (\ref{noncommchern}).
Observe that the  Gauss law of the Lagrangian (\ref{mcs-action1}) 
implies the following noncommutative condition on the matrices:
\ba
\left[ X_1,X_2 \right]=i \theta I,
\label{noncomm1}
\ea
with $I$ the identity matrix. This algebra only admits 
$\infty$-dimensional matrix representations: consider a particular, 
``ground state'' solution, $X=\hat{x}^i$, and write 
the general solutions as follows:
\ba
X^i=\hat{x}^i + \theta \epsilon^{ij} A_j(\hat{x}^i),
\label{pert}
\ea
where $A_i$ are $N \times N$ matrices of ``fluctuations'' 
($N \rightarrow \infty$). 
Note that these matrices can be expressed in terms of finite sums 
of products $e^{ip\hat{x}^1}e^{iq\hat{x}^2}$, i.e. they 
can be thought of being functions of $\hat{x}^i$. 
Replacing (\ref{pert}) in (\ref{mcs-action1}) and expressing 
the derivative as commutators, 
$\left[ \hat{x}^i, f(\hat{x}^1,\hat{x}^2) \right] = 
i \theta \epsilon^{ij} \partial_j f$, we find:
\ba
\mathcal{L} &=&\frac{\textbf{B}}{2} \Tr \left\{ 
-\theta \dot{A}_1 \left(\hat{x}_1 + \theta A_2 \right) + 
i A_0 \left[ \hat{x}_1 + \theta A_2, \hat{x}_2 - \theta A_1 \right] 
\right. \nl
&&
\qquad \left.
- \theta \dot{A}_2 \left( \hat{x}_2 - \theta A_1 \right) - 
 i A_0 \left[ \hat{x}_2 - \theta A_1, \hat{x}_1 + \theta A_2 \right] 
\right\} 
+ \textbf{B} \theta A_0  \nl \ 
&=& 
\frac{\textbf{B} \theta^2}{2} \Tr \left( A_1 \dot{A}_2 - \dot{A}_1 A_2 + 
2 A_0 (\partial_2 A_1 - \partial_1 A_2) + 2i A_0 \left[ A_1, A_2 \right] 
\right),
\label{lag-pert}
\ea
or in covariant notation,
\ba
\mathcal{L}=
\frac{\textbf{B}\theta^2}{2} \Tr \left[
- \epsilon^{\mu\nu\rho} A_\mu \partial_\nu A_\rho + 
\frac{2i}{3} \epsilon^{\mu\nu\rho} A_\mu A_\nu A_\rho \right]\ . 
\label{nc-cs}
\ea
In the limit $N \rightarrow \infty$ the matrix variables $A_i$ are mapped 
into smooth functions  of the noncommutative coordinates $A_i(\hat{x}_j)$. 
Also in this limit we can identify,
$\theta \Tr \rightarrow \frac{1}{2 \pi}\int d\hat{x}_1 d\hat{x}_2$,
and we obtain the following Lagrangian:
\ba
\mathcal{L}=
\frac{1}{4\pi \nu }\int d\hat{x}_1 d\hat{x}_2  
\left( - \epsilon^{\mu\nu\rho} A_\mu \partial_\nu A_\rho + 
\frac{2i}{3} \epsilon^{\mu\nu\rho} A_\mu A_\nu A_\rho \right). \
\label{nc-cs1}
\ea
In this Lagrangian, the fields $A_i$ still obey the matrix algebra, 
while in (\ref{noncommchern}) they are functions. 
The two formulations are related by expressing matrices as Wigner 
 $c$-number functions that obey the Moyal product algebra \cite{winf}.

Let us recall that any operator $\hat{F}(\hat{x_1},\hat{x_2})$ 
in the Weyl ordering $: \ :$ (most symmetric in $\hat{x_1},\hat{x_2}$) 
can be associated to a phase space function $F(x_1,x_2)$ as follows:
\ba
:\hat{F}(\hat{x_1},\hat{x_2}):
&=&
\int dx_1 dx_2\ F(x_1,x_2) : \delta(\hat{x_1} - x_1) \delta(\hat{x_2} - x_2): 
\nl
&=&  \int dx_1 dx_2 \frac{d\alpha}{2\pi} \frac{d\beta}{2\pi}\ F(x_1,x_2) 
e^{i\alpha (\hat{x_1} -x_1) + i \beta (\hat{x_2}-x_2)} \nl 
&=& F(-i \frac{\partial}{\partial \alpha}, -i\frac{\partial}{\partial \beta})
 e^{i\alpha \hat{x_1} + i\beta \hat{x_2}}  \vert _{\alpha=\beta=0} \ .
\label{opewig}
\ea
One finds by inspection that the product of two operators 
$\hat{F}$ and $\hat{G}$ corresponds to the Moyal product (\ref{moyal})
of the corresponding Wigner functions, 
$:\hat{F}:\ :\hat{G}:=:\hat{H}:$, $H(x_1,x_2)=(F \star G)(x_1,x_2)$.
In particular,
\ba
\int d \hat{x}_1 d \hat{x}_2 :\hat{F}(\hat{x}_1,\hat{x}_2):\ 
:\hat{G}(\hat{x}_1,\hat{x}_2): \ = \int dx_1 dx_2 (F \star G)(x_1,x_2) \ .
\label{noncomm3}
\ea
Therefore, the matrix Lagrangian (\ref{nc-cs1}) becomes 
the noncommutative Chern-Simons theory (\ref{noncommchern}) 
(within the Weyl ordering).

Finally, we recall that another route to obtain the Chern-Simons matrix 
theory (\ref{mcs-action1}), that emphasizes the discrete particle 
aspects of the fluid is given by a matrix regularization proposed 
by Goldstone and Hoppe \cite{hoppe}.

%-3.2------------------------------------------
\subsection{The Chern-Simons matrix model}

The noncommutative Chern-Simons Lagrangian (\ref{noncommchern}) 
and its matrix model formulation (\ref{mcs-action1}) both imply 
infinite degrees of freedom: therefore, Susskind's theory applies to an
infinite system. Instead, the fractional QHE is a system with a boundary and 
a finite number of particles.
Polychronakos introduced this feature \cite{poly1} by 
modifying Susskind's action (\ref{mcs-action1}) as follows:
\ba
S_{CSMM}=&& \int dt \frac{\textbf{B}}{2} Tr\left \{
\epsilon_{ab} (\dot{X}_a + i [A_0,X_a] ) X_b + 2 \theta A_0 - 
\s X_a^2 \right \} \nl 
&& + \int dt\ \psi^\dag(i\dot{\psi}-A_0 \psi). \
\label{CS-poly}
\ea
Two new terms are present: the first is a quadratic potential that 
confines the eigenvalues, i.e. keep the particles localized in the plane,
with $\s=O(\textbf{B}/N)$; 
the second term is a ``boundary'' $N$-dimensional complex vector $\psi$
that transforms in the fundamental representation of the gauge group $U(N)$.
The Gauss law is now given by:
\ba
G = \ -i \ {\textbf B} [ X_1 , X_2] + 
\psi \psi^\dag - {\textbf B} \theta I = 0. \
\label{poly-glaw}
\ea
Observe that the trace of (\ref{poly-glaw}) implies,
\ba
\psi^\dag \psi = N\textbf{B}\theta, \
\label{trace-glaw}
\ea
that can be realized with $N\times N$ dimensional matrices. 
The action (\ref{CS-poly}) thus defines the Chern-Simons matrix model, 
a gauge theory with $U(N)$ symmetry, 
$X_a \rightarrow U X_a U^\dag$, $\psi \rightarrow U \psi$ and 
$A_0 \rightarrow U A_0 U^\dag - i U \frac{d U^\dag}{dt}$.
Under a gauge transformation the action (\ref{CS-poly}) is not invariant, 
but it yields a winding number,
$ S \rightarrow S - i \textbf{B} \theta 
\int dt Tr \left[ U^\dag \dot{U}  \right]$; 
this requires the quantization
$\textbf{B} \theta=k $ \cite{nair}, leading to the 
Laughlin filling fractions (cf.(\ref{nu-class})).
Note that the equation of motion for $\psi$ in the $A_0=0$ gauge read, 
$\dot{\psi}=0$, showing that this is an auxiliary field with trivial 
dynamics; it can take the constant value 
$\psi=\sqrt{N \textbf{B} \theta} \mid v \rangle$, 
with $\mid v \rangle$ a vector of unit length \cite{poly1}.

%-3.3-----------------------------------------------
\subsection{Covariant quantization}

In the $A_0=0$ gauge, the Hamiltonian of the Polychronakos theory 
(\ref{CS-poly}) corresponds to $( N^2 + N )$ particles in the lowest 
Landau level with coordinates $X_{nm}$ and  $\psi_n$.  
It can be shown that, at quantum level, the Gauss law (\ref{poly-glaw}) 
implies gauge invariant states of the form 
$\Psi(X,\psi)=e^{-Tr(\bar{X}X)/2 - \psi^\dag \psi/2}\Phi(X,\psi)$, 
with $\Phi(X,\psi)$ a singlet of the gauge group $U(N)$ 
made by polynomials of $X_{nm}$ and $\psi_n$, being of order $Nk$ in $\psi_n$ 
due to (\ref{trace-glaw}) \cite{heller}.
A basis of states is given by:
\ba
\Phi(X,\psi)&=&\Phi_{ \{ n_1^1,...,n_N^1 \} }...\Phi_{ \{
n_1^k,...,n_N^k \} } \ ,\nl 
\Phi_{ \{ n_1^j,...,n_N^j \} } 
&=&
\epsilon^{i_1...i_N}(X^{n_1^j}\psi)_{i_1}...(X^{n_N^j}\psi)_{i_N}, 
\quad 0 \leq n_1^j < n_2^j < ... < n_N^j . \
\label{CS-states}
\ea
These states are eigenstates of the angular momentum $J$ with eigenvalues 
${\cal J}=N_X$, where $N_X$ is the number of matrices $X$ appearing in 
$\Phi(X,\psi)$. Since the Hamiltonian of the theory is proportional 
to the angular momentum, $H=(2\omega/{\bf B})\ J$,
 the states (\ref{CS-states}) are also eigenstates of the Hamiltonian.
The ground state of the theory is \cite{heller}:
\be
\Phi_{k-gs} = \left [ \epsilon^{i_1...i_N}
\psi_{i_1}(X\psi)_{i_2}...(X^{N-1}\psi)_{i_N} \right ]^k, \ 
\label{laughlin-wf}
\ee
corresponding to the lowest value of the angular momentum
(lower order polynomials vanish by 
antisymmetry of the $\epsilon^{i_1...i_N}$ tensor).

If we now diagonalize the matrix $X$ by the similarity transformation, 
$X = V^{-1} \Lambda V$, with $\Lambda=diag(\l_1,\dots,\l_N)$, we obtain:
\ba
\Phi_{k-gs}(V^{-1}\Lambda V,\psi) = C  \prod_{1 \leq n \leq m \leq N}
(\lambda_n-\lambda_m)^k .
\label{laughlin-wf-eigen}
\ea
The quantity $C$ appears in all the physical states and can
be neglected in the present discussion \cite{cr}.

Therefore we have obtained the Laughlin wave function as the ground
state of the Chern-Simons theory, with electron coordinates
identified with the eigenvalues of $X$. 
This is a very important result of the Chern-Simons matrix theory; 
that of reproducing the Laughlin wave function from gauge invariance 
of the states in presence of the ``background charge'' $\theta$.
(Note that the filling fraction is $\nu=1/(k+1)$: 
the shift from the classical value (\ref{nu-class}) is
due to a Vandermonde factor coming from the integration measure 
\cite{poly1}).

Let us now discuss the excitations over the ground state (\ref{laughlin-wf}).
In Ref.\cite{heller}, it was found the ``bosonic'' basis of states, 
\ba
\Phi \left(X,\psi\right) &=& \sum_{\{m_k\}}\ \Tr \left(X^{m_1}\right)\cdots
\Tr \left(X^{m_k}\right) \ \Phi_{k-gs}\ .
\label{HVR-bose}
\ea
for any positive integers $\{m_1,\dots,m_k\}$.
These states have $\D {\cal J} =r =\sum_k m_k$. 
For $r=O(1)$, their energy given by the boundary potential,
$\D E=\s \D {\cal J} = O(r\ \B/N)$ is very small:
they are the edge excitations of the droplet of fluid described 
by conformal field theories \cite{winf}\cite{wen}\cite{jainedge}.

The matrix model also possess localized density deformations 
that are analogues of the quasi-hole excitations of the Laughlin state. 
For example, the state
$\Phi_{\{n_1,\dots,n_N\}}$ in Eq.(\ref{CS-states}), with
$\{n_1,n_2,\cdots,n_M\}=\{1,2,\cdots,N\}$,
corresponds to moving one electron from the interior of the 
Fermi surface to the edge, causing $\D {\cal J} =O(N)$ and thus a finite gap
$\D E=O(\B)$.
On the other hand, the quasi-particle excitation cannot be
realized in the Chern-Simons matrix model \cite{poly1}\cite{cr1}\cite{cr2}, 
because excitations with angular momentum lower than (\ref{laughlin-wf}) 
are zero due to the antisymmetry of the $\epsilon^{i_1...i_N}$ tensor. 
Similarly, the Jain states $\nu=n/(2nk+1)$ are not naturally 
obtained \cite{poly3}.

In conclusion, we have shown that the Chern-Simons 
matrix model reproduces the Laughlin wave functions as ground states. 
Nevertheless, the theory has some drawbacks  \cite{hansson2}:
there are no quasi-particle excitations
\cite{poly1}, and the Jain states cannot be described \cite{poly3}.
Moreover, the measure of integration w.r.t. the eigenvalues $\l_i$
differs from that of electrons in the lowest Landau level, 
owing to the noncommutativity of matrices.
It can be shown that the ground state properties of the matrix theory and of
the Laughlin state only agree at long distances \cite{cr}\cite{karabali}.
Owing to the inherent noncommutativity, it is also difficult 
to match matrix observables
with  electron quantities of the quantum Hall effect \cite{hansson}.

%-3.4--------------------------------------
\subsection{Jain composite fermion transformation}

We would like to stress that the Chern-Simon matrix model
provides another realization of the Jain composite-fermion transformation
(see section 2.3). For $k=0$, the matrix theory reduced to the eigenvalues 
$\l_a$ is equivalent to a system of free fermions in the lowest Landau level,
i.e. to $\nu^*=1$ \cite{cr}\cite{karabali}. 
This fermionic picture is a general feature of one-dimensional 
matrix models \cite{mehta}.

In the presence of the $\th$ background, the noncommutativity
of matrix coordinates (\ref{poly-glaw}) forces the electrons 
to acquire a finite area of order $\th$, leading to the (semiclassical)
density $\rho_o=1/2\p\th$ (\ref{theta-def}).
Using this formula of the density and the quantization of $\B\th$,
we re-obtain the Jain relation (\ref{nu-star}) for flux attachment,
\be
\B\th \ = k\ \in\ \Z\ \ \to\ \ \B \ = \ k\ 2\pi\rho_o \ .
\label{delta-b-mcs}
\ee
Given that noncommutativity is expressed by the Gauss law of the matrix
theory, we understand that the mechanism for realizing
the Jain transformation is analogous to that of the Lopez-Fradkin theory 
(cf. (\ref{Fradfluid}), section 2.4),
but it is expressed in terms of different variables. 
However,  the higher Landau levels are not possible in the
Chern-Simons matrix model.

%-4--------------------------------------------
\section{Maxwell-Chern-Simons matrix gauge theory}

In this section we introduce and analyze the Maxwell-Chern-Simons 
matrix theory \cite{cr1} with the aim of improving the previous matrix
theory. The action is,
\ba
\!\!\!\! S&=&\int dt\ \Tr\left[
\frac{m}{2}\left(D_t\ X_i \right)^2  +
\frac{\B}{2} \eps_{ij} \ X_i\ D_t \ X_j  +
\frac{g}{2} \left[ X_1,X_2\right]^2 + \B\th\ A_0 \right]
\nl
& &-i\ \int \psi^\dag\ D_t \psi \ .
\label{mcs-action}
\ea
It involves the same $N\times N$ Hermitean matrices, 
$X_i(t)$ and $A_0(t)$, and the auxiliary vector $\psi(t)$, but 
includes two new terms with respect to Polychronakos theory (\ref{CS-poly}):
an additional kinetic term quadratic in time derivatives 
and a potential $V = - g \Tr \left[ X_1,X_2 \right]^2$, 
parameterized by the positive coupling constant $g$. 
All the terms in the action are fixed by the gauge principle 
because they are obtained by dimensional reduction of a gauge theory. 
Indeed, the action (\ref{mcs-action}) is the bosonic part of the low-energy 
effective theory of a stack of N D0-branes \cite{dzero1}
that has been discussed in the literature of string theory \cite{dzero}.  
In particular, D0-branes have been proposed as fundamental degrees 
of freedom in string theory \cite{mtheory}.

%-4.1---------------------------------------

\subsection{Low-energy effective action of 
Dp-branes in string theory}

Let us briefly review Witten's derivation of the effective low-energy action 
of $N$ Dp-branes \cite{witten}.
Consider ten-dimensional Minkowski space, with time 
$x^0$ and space $x^1, . . . , x^9$ coordinates, respectively.
A p-brane is an object that modifies the boundary conditions of open strings: 
it introduces Dirichlet boundary conditions in $(9-p)$ directions, as follows,
\ba
X^{p+1}(\sigma,t) 
&=& ....=X^{9}(\sigma,t)=0 \qquad \mbox{(Dirichlet)} , 
\label{bcond}
\\
\partial_{\sigma}X^1(\sigma,t) 
&=& .... = \partial_{\sigma}X^p (\sigma,t)=0
\quad \mbox{(Neumann)} .
\ea
Due to (\ref{bcond}), the zero modes $X^j$ with $j>p$ are frozen, 
and the massless particles are functions of $X_1,..,X_p$ only. 
The massless bosons $A_i(X^s)$, $i,s=0,...,p$, propagate as $U(1)$ 
gauge bosons on the p-brane, while the other components become 
scalars fields on the p-brane, $\phi_j(X^s)$, $j>p$.
The vertex operators for insertions of spin-$1$ fields in string theory 
are given by:
\ba V_A &=& \sum_{i=0}^p A_i(X^s)\partial_\tau X^i, \nl 
V_\phi &=& \sum_{j>p} \phi_j(X^s)
\partial_\sigma X^j . 
\label{vert-op} \ea
For $\phi_j=$constant, the boundary integral of $V_\phi$ implies 
the change $X^j \rightarrow X^j + \phi^j$ for $j>p$: thus the scalars
$\phi_j,j>p$ can be interpreted as the coordinates  of the p-brane.
The theory on the $(p + 1)$ dimensional brane world-volume is 
naturally the ten-dimensional $U(1)$ supersymmetric gauge theory
dimensionally reduced to $(p + 1)$ dimensions.

Bound states of $N$ parallel Dirichlet p-branes can be described 
by the low-energy limit when the branes are nearby. 
We consider the case of two parallel Dirichlet p-branes, 
one at $X^j = 0$, and one at $X^j = a^j$ ($j>p$). 
The branes are connected by strings: they can start and end on the 
same brane and give a $U(1) \times U(1)$ gauge theory 
(with one $U(1)$ living one each p-brane), or they can start
in the first brane and end in the second (and viceversa). 
In this case, the strings have $U(1) \times U(1)$ charges. 
The ground state of this
configuration has an energy $T \mid a \mid$, with $T$ and $\mid a \mid$ 
being the tension and length of the string, respectively. 
When $\mid a \mid \rightarrow 0$
the charged vector bosons become massless and the $U(1) \times U(1)$ 
gauge symmetry is enlarged to a $U(2)$ symmetry.
In the same way, $N$ coincident parallel branes yield 
a $U(N)$ gauge symmetry on the p-brane.
The field content in the effective action is given by 
the $U(N)$ gauge field $A_j(X^s,t)$, with $ s,j=1,...,p$ , 
and the scalar fields $\phi_j(X^s,t)$, with$ j>p$ , 
in the adjoint representation of $U(N)$, i.e. they are all 
$N \times N$ matrices.

The reduction to $(p+1)$ dimensions of the bosonic sector of 
the theory is obtained as follows. From the Lagrangian,
\ba
L_{YM}=-\frac{1}{4g^2} Tr \left( F^{\mu\nu} F_{\mu\nu} \right)\ ,
\label{usual-YM}
\ea
we simplify the commutators in $F^{\mu\nu}= \left[ D_{\mu},D_{\nu} \right] $
by dimensional reduction and identify the earlier fields, leading to:
\be
L'_{YM} =
- \frac{1}{4 g^2}Tr \left( \sum_{r,s=0}^p F^{rs} F_{rs} 
- \sum_{s=1}^p\sum_{j>p} D_s\phi_j D^s\phi^j 
+\ \sum_{i,j>p} \left[ \phi_i, \phi_j \right]^2 \right).
\label{dim-red-YM}
\ee
In the $p=0$ case we have D0-branes, that are nonrelativistic point particle
with matrix variables and one-dimensional gauge symmetry. 
In (2+1)-dimensions, (\ref{dim-red-YM}) becomes 
the Maxwell-Chern-Simons theory (\ref{mcs-action}), apart from 
the Chern-Simons kinetic term: as shown in Ref.\cite{dzero}, this
can be obtained by adding a configuration of higher D-branes 
that creates a magnetic field for the D0-branes.

%-4.2---------------------------------------------------------

\subsection{Covariant quantization of Maxwell-Chern-Simons theory}

In this section we quantize the 
Maxwell-Chern-Simons matrix theory (\ref{mcs-action}) \cite{cr1}\cite{park}.
The canonical momenta are given by the Hermitean matrices:
\be
\P_i = \frac{\d S}{\d \dot{X}^T_i}\ = \
D_t X_i -\frac{\B}{2} \eps_{ij} X_j \ ,
\label{pi-def}
\ee
and  the vector $\chi=\d S/\d \dot{\psi}= -i\psi^\dag$. 
The Hamiltonian is:
\be
H \ =\ \Tr \left[
\frac{1}{2}\left(\P_i +\frac{\B}{2}\ \eps_{ij}\ X_j \right)^2 \ -\
\frac{g}{2} \left[X_1,X_2 \right]^2\
\right]\ .
\label{mcs-ham}
\ee
The Gauss law constraint now reads:
\be
G\ = \ 0\ , \qquad
G\ =\ i\ \left[X_1,\P_1\right]\ +\ i\ \left[X_2,\P_2\right]\
-\ \B\th\ I \ +\ \psi\otimes \psi^\dag \ .
\label{mcs-gauss}
\ee
As in Chern-Simons theory, 
G generates $U(N)$ gauge transformations on $X_i$ and $\psi$ at the 
quantum level, and requires the physical states to be $U(N)$ singlets.
We now quantize all the $2N^2$ matrix degrees of freedom $X^i_{ab}$ and
later impose the Gauss law as a differential condition on wave functions.
It is useful to introduce holomorphic coordinates:
\ba
X &=& X_1+i\ X_2\ , \qquad\qquad \ov{X}=X_1 -i\ X_2\ ,
\nl
\P &=& \frac{1}{2}\left(\P_1 -i\ \P_2\right)\ ,
\qquad \ov{\P}=\frac{1}{2}\left(\P_1 +i\ \P_2\right)\ ,
\label{b-mat}
\ea
with the bar denoting the Hermitean conjugate of
classical matrices, keeping the dagger for the quantum adjoint.

The Hamiltonian (\ref{mcs-ham}) for $g=0$ is quadratic and easily solvable:
 introduce the matrix,
\be
A_{ab} = \frac{1}{2\ell}  X_{ab} + i\ell\  \ov{\P}_{ab}\ ,
\label{a-def}
\ee
and its adjoint $A^\dag$. Owing to 
the canonical commutators, they obey the algebra of $N^2$ harmonic oscillators:
\be
 \left[\left[A_{ab},A^\dag_{cd} \right]\right] \ = \ \d_{ad}\ \d_{bc}\ \ ,
\qquad\qquad
\left[\left[A_{ab},A_{cd}\right]\right] \ =\  0\ .
\label{a-comm}
\ee
In this following, the double brackets describe quantum commutators
while the single ones are kept for the matrix algebra;
note also that $A^\dag$ is the adjoint of $A$ both as a matrix and a
quantum operator. The Hamiltonian can be expressed in term of 
$A$ and $A^\dag$ as follows:
\be
H \ =\ \B\ \Tr \left( A^\dag \ A \right)\ +\ \frac{\B}{2} N^2\ +
\frac{g}{8} \Tr \left[ \bar{X},X \right]^2\  .
\label{a-ham}
\ee
In the term $\Tr(A^\dag A)=\sum_{ab} A^\dag_{ab} A_{ba} $ one recognizes
$N^2$ copies of the Landau level Hamiltonian corresponding
to $N^2$ two-dimensional ``particles'' with phase-space coordinates,
$\{\P_{ab},X_{ab}\}$ and $\{\bar{\P}_{ab},\bar{X}_{ab}\}$, $a,b=1,\dots,N$.
The one-particle state are similarly characterized by another set of
independent oscillators corresponding to angular momentum excitations,
that are described by the matrix $B=\bar{X}/2\ell + i\ell\  \P$
and its adjoint $B^\dag$, obeying the algebra:
\be
 \left[\left[B_{ab},B^\dag_{cd} \right]\right] \ = \ \d_{ad}\ \d_{bc}\ \ ,
\qquad\qquad
\left[\left[B_{ab},B_{cd}\right]\right] \ =\  0\ .
\label{b-comm}
\ee
The total angular momentum of the $N^2$ ``particles'' can be
written in the $U(N)$ invariant form,
\be
J\ =\ \frac{1}{i} \Tr \left( \bar{X} \bar{\P} -  X \P  \right)\ = \
 \Tr\left( B^\dag B\ - \ A^\dag A\right)\ .
\label{j-def}
\ee

For large values of the magnetic field $\B$, the reduction of
the theory (\ref{mcs-action}) to the lowest Landau is obtained by  
imposing $A_{ab}\ap 0$, $ \ \forall a,b$: the theory becomes the
previously studied Chern-Simons matrix model (\ref{CS-poly}), because
the quadratic kinetic term vanishes and the potential 
reduces to a constant due to the Gauss law.

%-5--------------------------------------------------

\section{Matrix ground states at $g=0$}

\subsection{Jain states  by projections}

 The gauge invariant states can be written,
\be
\Psi\ = \ e^{-\Tr\left(\ov{X} X \right)/2 -\ov{\psi} \psi/2}\
\Phi(X,\ov{X},\psi) \ ,
\label{wf-def}
\ee
where $\Phi(X,\ov{X},\psi)$ is again a $U(N)$ singlet
made of matrices $X$, $\ov{X}$ and $Nk$ components of the vector $\psi$. 
The general solutions (\ref{wf-def}) are similar to those obtained 
in Chern-Simons theory with the difference that now the polynomial part
also depends on the $\ov{X}$ matrices; for example,
\be
\Phi(X,\ov{X},\psi)=\left(\epsilon^{i_1 i_2 .... i_n} 
\psi_{i_1}( X \ov{X} \psi )_{i_2} .... 
(X \ov{X} X \ov{X} X ...\psi)_{i_n} \right)^k .
\label{state-max}
\ee
It is better to express these polynomials in terms of 
the variables $\ov{A}$ and $\ov{B}$ (cf. section 4.2):
from the commutation relations (\ref{a-comm},\ref{b-comm}), the
energy and momentum eigenstates can be easily obtained by applying the
$A^\dagger_{ab}$ (\ref{a-def}) and $B^\dagger_{ab}$ operators
to the empty ground state 
$\Psi_o=\exp\left(-\Tr \ov{X}X/2 - \ov{\psi}\psi/2\right)$.
Their energy  $E=\B N_A$ and momentum ${\cal J}=N_B-N_A$ are expressed 
in terms of the number
of $A^\dag$ and $B^\dag$ operators, $N_A$ and $N_B$ respectively.
The wave functions is rewritten:
\be
\Psi\ = \ e^{-\Tr\ \ov{X} X/2 -\ov{\psi} \psi/2}\
\Phi(\ov{B},\ov{A},\psi) \ ,\quad
E=\B\ N_A\ , \quad J=N_B-N_A\ ,
\label{wf-def2}
\ee
 where
$\ov{B}=X-\de/\de\ov{X}$ and  $\ov{A}=\ov{X}-\de/\de X$ commute
among themselves, $[[\ov{A}_{ab},\ov{B}_{cd}]]=0$,
and can be treated as $c$-number matrices. Any polynomial 
$\Phi(\ov{B},\ov{A},\psi)$ yields an energy eigenstate and corresponds
in general to a sum of terms (\ref{state-max}).
Let us remark that for states with constant density, 
the angular momentum measures 
the extension of the ``droplet of fluid'',  such that we can associate 
a corresponding filling fraction $\nu$ by the formula,
\be
\nu\ =\ \lim_{N\to\infty}\ \frac{N(N-1)}{2{\cal J}}\ .
\label{nu-def}
\ee

\begin{figure}
\begin{center}
\includegraphics[width=8cm]{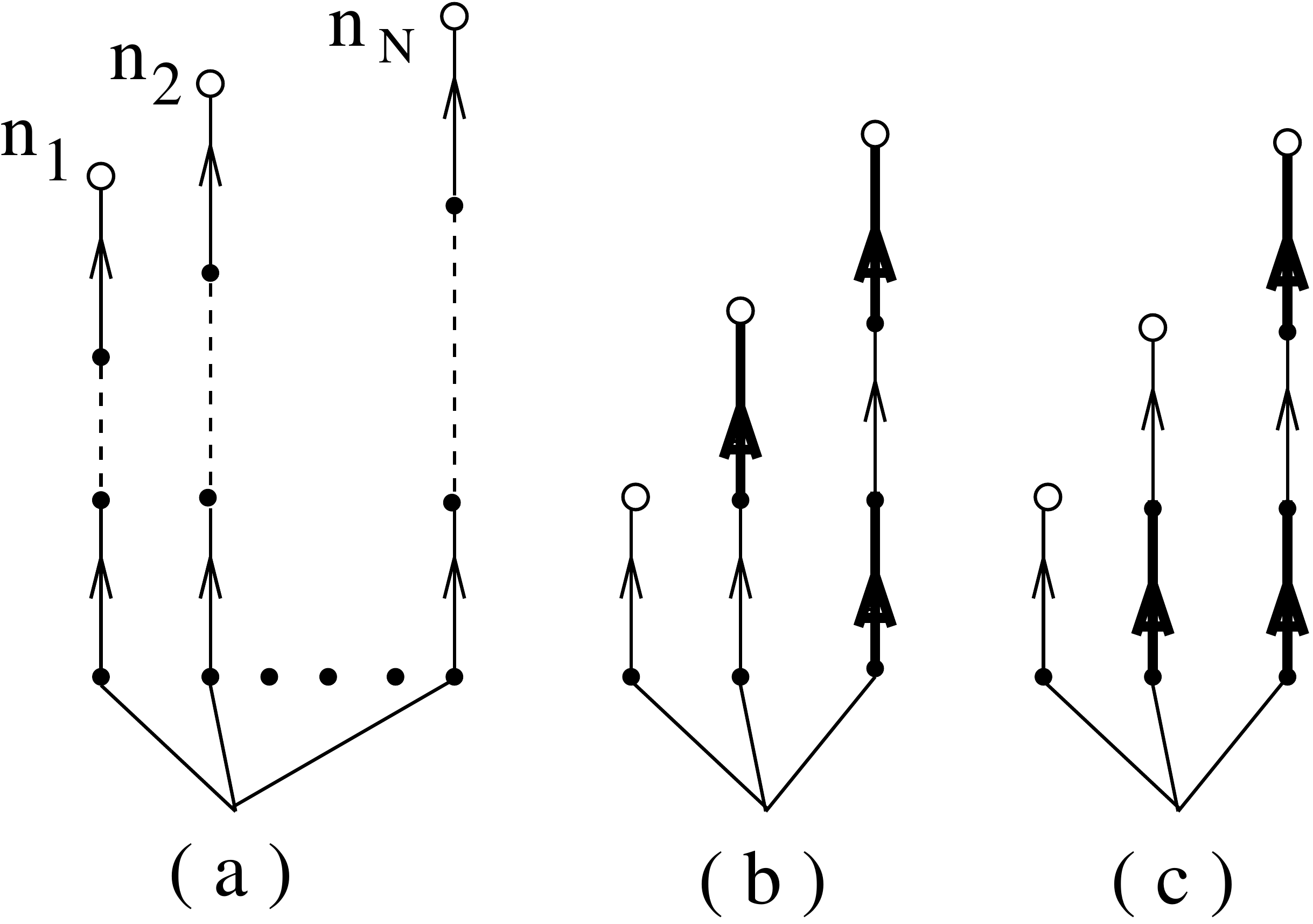}
\end{center}
\caption{Graphical representation of gauge invariant states: (a) 
general states in the lowest Landau level (cf. Eq.(\ref{CS-states})); 
(b) and (c), $N=3$ examples involving both
matrices, $\ov{B}$ (thin line) and $\ov{A}$ (in bold).}
\label{bush}
\end{figure}

The states (\ref{wf-def2})
can be represented graphically as ``bushes'', as shown in Fig.\ref{bush}(a).
The matrices $\ov{B}_{ab}$ (i.e. $X_{ab}$) are depicted as oriented 
segments with indices at 
their ends and index summation amounts to joining segments
into lines, as customary in gauge theories.
The matrices $\ov{A}_{ab}$ are represented by bold segments.
The lines are the ``stems'' of the bush ending with one $\psi_a$, represented
by an open dot, and the epsilon tensor is the N-vertex
located at the root of the bush. 
Bushes have N stems  of different lengths: $n_1 < n_2 <\cdots <n_N$. 
The position $i_\ell$ of one
$\ov{B}$ on the $\ell$-th stem, $1\le i_\ell\le n_\ell$, 
is called the ``height'' on the stem. 
Since two stems cannot be equal, one obtains a kind of Fermi sea 
of $N$ ``one-particle states'' corresponding to the $N$ strands.
However,  there are additional degeneracies with 
respect to an ordinary fermionic system, because in each stem all possible
words of $\ov{A}$ and $\ov{B}$ of given length yield independent states
(for large $N$), owing to matrix noncommutativity, 
as seen in fig.\ref{bush}(b) and \ref{bush}(c).

%-f4------------------------------------------
 
The complete filling of all the available degenerate $E>0$ states at $g=0$
clearly gives very dense and inhomogeneous fluids that are incompatible
with the physics of the quantum Hall effect. The matrix degeneracies
lead to a  density of states that grows exponentially with the energy;
this is a characteristic of string theories that is
not suitable for the Hall effect \cite{mtheory}.
On the other hand, for $g>0$ the potential $\Tr [X,\ov{X}]^2$ 
in the Hamiltonian (\ref{a-ham}) constraints
matrix noncommutativity and eventually eliminates the degeneracy: 
at $g=\infty$, this is not present and the theory can describe
a physical electron system, as shown in section 6.

Given that the $g>0$ theory is difficult to solve, in ref.   
\cite{cr1} we introduced  a set of projections that
limit the matrix degeneracy at $g=0$ and are explicitly solvable.
These projections are expressed by the following 
constraints on the wave function,
\be
\left(A_{ab}\right)^m\ \Psi\ = \ 0\ \longrightarrow\ 
\left(\frac{\de\ \ }{\de \ov{A}_{ab}}\right)^m
\ \Phi(\ov{A},\ov{B},\psi) \ =\ 0\ ,\qquad
\forall\ a,b \ ,
\label{sll-cond}
\ee
for a given value of $m$. The $m=1$ case is the lowest Landau level
discussed before with no $\ov{A}$ dependence, while $m$ taking 
successive values $m=2,3,\dots$ gradually allow larger $\ov{A}$ 
multiplicities and thus matrix degeneracies.
Note that in equation (\ref{sll-cond}), each matrix component 
$A_{ab}$ is raised to the $m$-th power,  without index summation:
the condition is nevertheless gauge invariant and admits an equivalent
manifestly invariant form that is discussed later.

The results of ref.\cite{cr1} were rather interesting: not only 
the projections (\ref{sll-cond}) allow homogeneous ground states suitable 
for describing quantum Hall fluids, but also they precisely occur in
the Jain pattern of filling fractions, $\nu=m/(mk+1)$, and their
derivation repeats step-by step the Jain ``composite fermion''
construction \cite{jain}.
 
Let us recall the main points of the analysis. 
Consider first the projection (\ref{sll-cond}) for $m=2$ and choose $k=1$: 
the solutions are polynomials that are at most linear
in each component $\ov{A}_{ab}$.
Let us imagine that one or more $\ov{A}$ matrices are present at points on 
the bush as in Fig.(\ref{bush}).
The differential operator  (\ref{sll-cond})
acts by sequentially erasing pairs of bold lines on the bush,
each time detaching
two strands and leaving four free extrema with indices fixed to 
either $a$ or $b$, with no summation on them.
For example, when acting on a pair of $\ov{A}$ located on the same stem, 
it yields a non-vanishing result: this limits the bushes to have 
one $\ov{A}$ per stem at most.
The $A^2 \ap 0$ conditions can be satisfied if cancellations occur 
for pairs of $\ov{A}$ on different stems, as it follows:
\be
\left(A_a^b\right)^2\Phi =\cdots \ +\  \eps^{\dots i \dots j \dots}
\left(
\cdots M_{ia}\ N_{ja}\cdots V^b\ W^b \right)
\ +\ \cdots\ ,
\quad (a,b\ {\rm fixed}).
\label{sll-canc}
\ee
This expression vanishes for $M=N$ due to the
antisymmetry of the epsilon tensor.
The general solution of (\ref{sll-cond}) is given by bushes involving one
$\ov{A}$ per stem at most (max N matrices in total), with all of
them located at the same height on the stems \cite{cr1}. In formulas:
\ba
\Phi_{\{n_1,\dots,n_\ell; p; n_{\ell+1},\dots,n_M\}} 
&=& \eps^{i_1\dots i_N} \ 
\prod_{k=1}^{\ell}\ 
\left(\ov{B}^{n_k}\psi\right)_{i_k} \
 \prod_{k=\ell+1}^{N}\ 
\left(\ov{B}^p\ov{A}\ \ov{B}^{n_k} \psi\right)_{i_k}\ , 
\nl
&&\quad 0\le n_1 < \cdots < n_\ell\ ,\quad
0\le n_{\ell+1} < \cdots < n_N\ . 
\label{two-ll}
\ea

If the matrices $\ov{A},\ov{B}$ were diagonal, 
these states would be Slater determinants of ordinary Landau levels.
The matrix states have further degeneracies by commuting $\ov{A},\ov{B}$
pairs: however, commutations are severely limited
in the the solution (\ref{two-ll}), only the global $p$ dependence
is allowed. This shows how the $A^2 \ap 0$
projection works in reducing matrix degeneracies.

The ground state in the $A^2 \ap 0$ theory with finite-box boundary 
conditions is the lowest $J$ states of the form (\ref{two-ll}):
it corresponds to {\it homogeneous} filling 
all the allowed states in the first and second Landau levels with $N/2$
``gauge invariant particles'' each, and reads:
\be
\Phi_{1/2, \ gs} = \eps^{i_1\dots i_N} \ 
\prod_{k=1}^{N/2}\ 
\left(\ov{B}^{k-1}\psi\right)_{i_k} \ 
\prod_{k=1}^{N/2}\ 
\left(\ov{A}\ \ov{B}^{k-1} \psi\right)_{i_{N/2+k}}\ , 
\label{two-jain}
\ee
with angular momentum ${\cal J}=N(N-4)/4$.
This state is non-degenerate due to the vanishing of the 
$p$ parameter in (\ref{two-ll}). It has filling fraction
$\nu^*=2$, assuming homogeneity of its density, to be shown later.

The ground states for $k>1$ are products of $k$
bushes: they obey the constraint $A^2\ap 0$ provided that 
the two derivatives always vanish when distributed over the bushes.
Given one bush of type (\ref{two-jain}),  
obeying $A^2\ \Phi_{1/2, \ gs}=0$, one can form the state,
\be
\Phi_{k+1/2, \ gs} \ = \ \Phi_{k-1, \ gs} \ \Phi_{1/2, \ gs} \ ,
\label{two-k-jain}
\ee
where the other $(k-1)$ bushes 
satisfy $A\ \Phi_{k-1,\ gs} =0$ and actually are Laughlin's one
(\ref{CS-states}).
Using (\ref{nu-def}), the angular momentum of this state corresponds to the
filling fraction $1/\nu=k+1/2$.

We thus find the important result that the $A^2\ap 0$ projected
Maxwell-Chern-Simons theory possesses non-degenerate ground states 
that are the matrix analogues of the Jain states  obtained by 
composite-fermion transformation at $\nu^*=2$, 
i.e. $1/\nu=1/\nu^* + k$.
The matrix states (\ref{two-k-jain},\ref{two-jain}) 
would actually be exactly equal to Jain's wave functions, if the
$\ov{A},\ov{B}$ matrices were diagonal: the $\psi$ dependence
would factorize and the matrix states 
reduce to the Slater determinants of the Jain wave functions
\cite{jain}\cite{numeric}
(cf. (\ref{jain-elecstate}) in section 2.3).

The correspondence extends to the whole Jain series:
the other $\nu^*=m$ non-degenerate ground states
 are respectively obtained in the theories with $A^m\ap 0$ projections.
They read:
\be
\Phi_{k+1/m, \ gs}\ =\  \Phi_{k-1, \ gs} \ \Phi_{1/m, \ gs} \ ,
\label{m-k-jain}
\ee
where,
\be
\hspace{-2.5cm} \Phi_{1/m, \ gs} =  \eps^{i_1\dots i_N} \ 
\prod_{k=1}^{N/m}\ \left[
\left(\ov{B}^{k-1}\psi\right)_{i_k} \ 
\left(\ov{A}\ \ov{B}^{k-1} \psi\right)_{i_{k+N/m}}
\cdots
\left(\ov{A}^{m-1} \ \ov{B}^{k-1} \psi\right)_{i_{k+(m-1)N/m}}
\right].
\label{m-jain}
\ee

In conclusion, in ref.\cite{cr1} we found that the ground states 
of the properly projected Maxwell-Chern-Simons matrix theory 
reproduce the Jain pattern of the composite fermion construction \cite{jain};
the matrix states are non-degenerate for specific values of the
density that are controlled by the boundary potential \cite{cr1}.
These results indicate that the Jain composite fermions
have some relations with the D0-brane degrees of freedom
and their underlying gauge invariance. 
Both of them have been described as dipoles: according to Jain
\cite{jain} and Haldane-Pasquier \cite{pasquier}, 
the composite fermion can be considered as the bound state
of an electron and a hole (a vortex in the electron fluid).
On the other side, matrix gauge theories, and the equivalent noncommutative
theories \cite{mtheory}, describe D0 branes that are point-like dipoles 
in the low-energy limit of string theory.

A final remark on the noncommutative matrix coordinates in the Jain and 
Laughlin state: the Gauss law (\ref{mcs-gauss})
can be rewritten in terms of $X,\ov{X},A,\ov{A}$ as follows:
\be
\left[X,\ov{X}\right] \ +\ \frac{2}{B}
\left[\ov{X},A\right] \ +\  \frac{2}{B} \left[\ov{A},X\right] \ = \
2\left( \theta - \frac{1}{B}\psi\otimes\ov{\psi} \right)\ .
\label{not-non-com}
\ee
For the Laughlin states in the lowest Landau level,
this reduces to coordinates noncommutativity (\ref{poly-glaw}),
because $A=\ov{A}=0$; for the Jain states populating higher levels, there are
other terms contributing to noncommutativity besides the matrix
coordinates, such that higher density values are possible.

%-5.3------------------------

\subsection{Gauge invariant form of the projections}

Although the operators $(A_{ab})^m$, $ m=1,2,..$,  
are not explicitly gauge invariant, their kernel restricted to gauge invariant 
states yields gauge invariant conditions, as seen in the previous discussion.
Therefore, the projectors should have manifestly gauge invariant expressions. 
In Ref. \cite{cr2}, they were found by expressing the conditions
$A^m\ap 0$ in terms
of positive-definite occupation numbers $Z_{ab}=A^\dag_{ab} A_{ab}$
(no sum over $a,b$), and by averaging over their gauge orbit.
For $m=2$, the $A^2\ap 0$ constraint was shown to be 
equivalent to $Q_2^{g.i.}\ap 0$, with:
\be
Q_2^{g.i.}\ \propto \ \left(\d_{ki}\ \d_{lj}\ +\ \d_{kj}\ \d_{li}\right)\ 
A^\dag_{ia'}\ A^\dag_{ja} \ A_{ak}\ A_{a'l}\ . 
\label{q2-gi}
\ee
Upon commuting the operators to bring summed indices close to each other, we
finally find the manifestly gauge-invariant form 
(disregarding the normalization):
\be
Q_2^{g.i.}\ = \ \Tr\left(A^\dag A A^\dag A\right)\ 
+\ \left( \Tr\ A^\dag A \right)^2\ -\ (N+1)\ \Tr\left( A^\dag A\right)\ .
\label{q2-gi-fin}
\ee
One can check that the action of the gauge-invariant constraint 
$Q^{g.i.}_2$ on bush wave functions is completely equivalent 
to that of the gauge-variant condition $A^2\ap 0$ \cite{cr1}.
The expressions (\ref{q2-gi}) easily generalizes to higher $m$ values
\cite{cr2}.

%-5.4------------------------------------------------
\subsection{Semiclassical solutions at $g=0$}

In this section we review the semiclassical analysis of the $g=0$ 
 Maxwell-Chern-Simons theory: in Ref. \cite{cr2},
we found the semiclassical states that correspond to the quantum states
with homogeneous filling and composite-fermion structure (\ref{m-k-jain})
of the previous section and some of their quasiparticle excitations.
The motivations for the semiclassical analysis are twofold: on one side,
previous experience \cite{poly1}\cite{hansson2}\cite{hansson}
\cite{mtheory}\cite{lambert} with noncommutative field theory 
has shown that the classical fluid dynamics incorporates some properties
of the full quantum theory. From another side, it is know that the
Laughlin states in the quantum Hall effect are incompressible fluids
that become semiclassical in the thermodynamic limit $N\to\infty$
\cite{winf}. 

As we showed in section 5.1, the Jain-like ground states (\ref{m-k-jain})
involve higher Landau levels ($A \neq 0$) and have
filling fractions  $\nu^*=2,3,\dots$  (cf. (\ref{nu-star}) in section 2.3).
We first note that these states are characterized by
 $E=O(N)$ and $J=O(N^2)$, thus implying that
the matrix $A$ must have few nonvanishing elements $O(1)$.
Indeed, the constraint $A^m \ap 0$ can be written in terms of 
occupation numbers, $Z_{ab}=\ov{A}_{ab} A_{ab}$, and limit
the semiclassical values of $A_{ab}$ matrix elements: 
once summed over each row or column, they can take the values 
$\g=0,1,\dots,m-1$ \cite{cr2}.

We introduce the constraints of the Gauss law and the projection 
$A^m\ap 0$ in the Hamiltonian with Lagrange multipliers $\L$ and 
$\G_a,\G'_b$, respectively.
Upon variation with respect to $\ov{A},\ov{B}$, 
we obtain the equations of motion:
\ba
i\ \dot{A}_{ab} &=& 2 A_{ab}\ -\ \left[ \Lambda , A \right]_{ab}\ 
+\ A_{ab}\left( \G_a \ +\ \G'_b \right)\ ,
\nl
i\ \dot{B} &=& - \left[ \Lambda, B \right]\ + \w \ B \ ,
\nl
G &=& \left[ \ \ov{A} , A \ \right]\ +\ 
\left[ \ \ov{B} , B \ \right]\ - k + \psi\otimes \ov{\psi} = 0 \ ,
\nl
Z_a &=& \sum_b\ \ov{A}_{ba}\ A_{ab}\ =\ \g\ , 
\qquad\qquad\qquad \g=0,1,\dots, m-1\ ,
\nl
Z'_b &=& \sum_a\ \ov{A}_{ba}\ A_{ab}\ =\ \g'\ , 
\qquad\qquad\qquad \g'=0,1,\dots, m-1\ .
\label{eq-mot}
\ea
The semiclassical ground states correspond to solutions with
$\dot{A}=\dot{B}=0$.

Let us first recall the classical ground state with $A_{ab}\ap 0$ 
(lowest Landau level) found by Polychronakos in the Chern-Simons matrix model 
 \cite{poly1}: in this case ($\ell=1$),  $\ov{B}=X$ and $\psi$ are given by,
\ba
\bar{X} \ = \ \sqrt{k} \
\sum_{n=1}^{N-1} \sqrt{n} \vert n\ra \la n - 1 \vert \ ,\qquad
\psi =\sqrt{kN} \vert N-1\ra\ .
\label{sol-poly}
\ea
(denoting $|0\ra,\cdots,|N-1\ra$ the basis vectors \cite{cr2}).
The radius-squared matrix coordinate $R^2$ is diagonal, and given by:
\ba
R^2 \ = \ov{X} X \ = {\rm diag}\left(0,k,2k,\dots,(N-1)k \right)\ .
\label{polyr1}
\ea
From the distribution of the eigenvalues in (\ref{polyr1}), 
it is clear that this solution implies an uniform density. 
In the large N limit, the filling fraction takes the Laughlin values 
$\nu=1/k$ according to the formula (\ref{nu-def}).

In general, the one-particle density of rotation invariant states
in matrix models can be defined in terms of the
gauge invariant eigenvalues of $R^2$, as follows ($\rho(r)= \rho(r^2)/ \pi$):
\ba
\rho(r^2) = \sum_{i=0}^{N-1} \delta(r^2 - \sigma_i), \qquad \sigma_i
\in Spec(R^2).
\label{cont-dens}
\ea
For semiclassical fluids, this becomes a piecewise continuous function 
in the limit $N \to \infty$. A discrete approximation suitable 
for the continuum limit is \cite{cr2}:
\ba
\rho(r^2) = \sum_i \frac{n_i}{\sigma_{i+1} - \sigma_i} \delta_{r^2,\sigma_i}
\ ,
\label{discrete-dens}
\ea
involving the Kronecker delta and the ordered set of distinct eigenvalues, 
$\sigma_i<\sigma_j$, $i<j$, with multiplicities $n_i$.

From Ref.\cite{poly1}, we also recall the form of the quasi-hole 
in the origin, in the lowest Landau level:
\ba
\bar{X} = \sqrt{k} \left( \sqrt{q} \mid 0 \rangle \langle N - 1
\mid + \sum_{n=1}^{N-1} \sqrt{n+q} \mid n \rangle \langle n - 1 \mid \right),
 \quad q>0,
\label{quasip}
\ea
where $q$ is proportional to the charge of the quasi-hole.
The $R^2$  eigenvalues are correspondingly shifted
upward by $q$, causing a dip at the origin.

The semiclassical ground state solution for $A^2\ap 0$, leading to 
the Jain like $\nu^*=2$ ground state is found by a suitable
ansatz to the equation of motions (\ref{eq-mot}). 
After gauge choice, they imply that: i) $B$ is again a raising
operator as in (\ref{sol-poly}) and ii) $A$ has only one element $1$ in each
row and column, i.e. it is a (partial) permutation matrix.
Using these data, the matrix equations can be reduced to a linear 
system of $O(N/2)$ conditions leading to the solution ($N$ even):
\ba
B &=& \sum_{n=1}^{N/2} \sqrt{n(k-1)} \mid n \ra \la n-1 \mid\ + \
\sum_{n=\frac{N}{2}+1}^{N-1} \sqrt{ n(k+1)-N} \mid n \ra \la n-1 \mid \ ,
\nl
A &=& \sum_{n=0}^{\frac{N}{2}-1} \mid n+ \frac{N}{2} \ra \la n \mid \ ,
\label{nu2-sol}
\ea
with $\psi$ as in (\ref{sol-poly}).
In matrix form for $N=4$, it reads:
\be
\!\!\!\!\!\!\!\!\!\! B = \left(\begin{array}{cccc}
0 & 0 & 0 & 0 \\ \sqrt{k-1} & 0 &  0 & 0 \\
0 & \sqrt{2(k-1)} & 0 & 0 \\ 0 & 0 & \sqrt{3k-1} & 0
\end{array}\right) \ , \quad
A = \left(\begin{array}{cccc}
0 & 0 &  0 & 0 \\ 0 & 0 & 0 & 0 \\ 1 & 0 & 0 & 0 \\
0 & 1 & 0 & 0 \end{array}\right) . \
\label{nu2-mat}
\ee
This solution has same energy $E=\B N/2$ of the quantum state
(\ref{two-k-jain}) and same angular momentum $J=(k-1/2)N^2/2+O(N)$  to leading
order. The matrix $R^2=(B+\ov{A})(\ov{B}+A)$ contains off-diagonal terms
from the mixed products: however, these give subdominant $O(1/\sqrt{N})$
corrections to the eigenvalues as is clear in a simple two-by-two matrix
example. Thus, ${\rm Spec}(R^2)= {\rm Spec}(B\ov{B})(1+O(1/\sqrt{N}))$.

%-f5-----------------------------------
\begin{figure}
\begin{center}
\includegraphics[width=7cm]{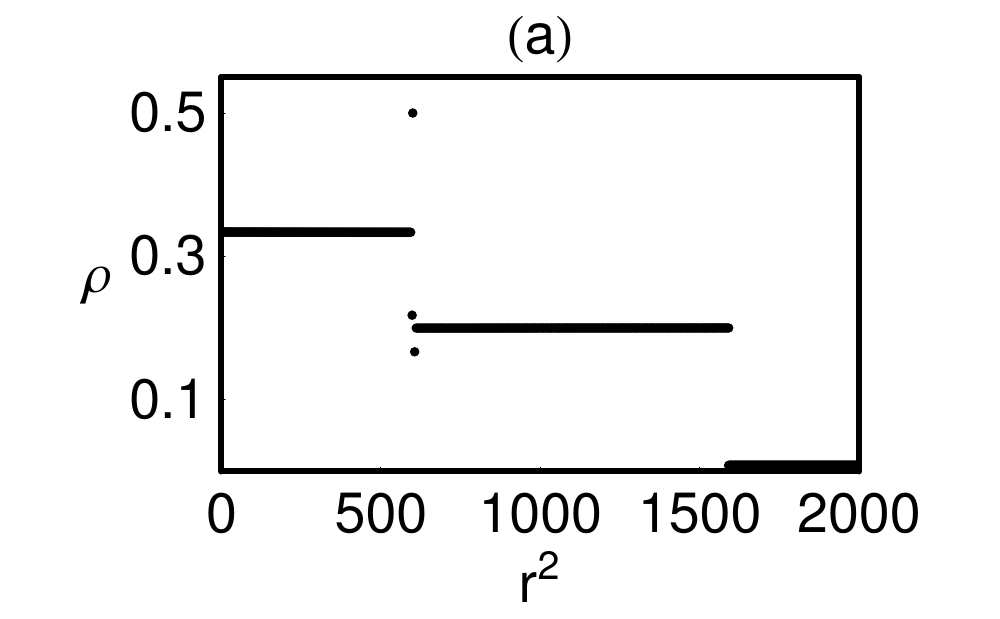}
\includegraphics[width=7cm]{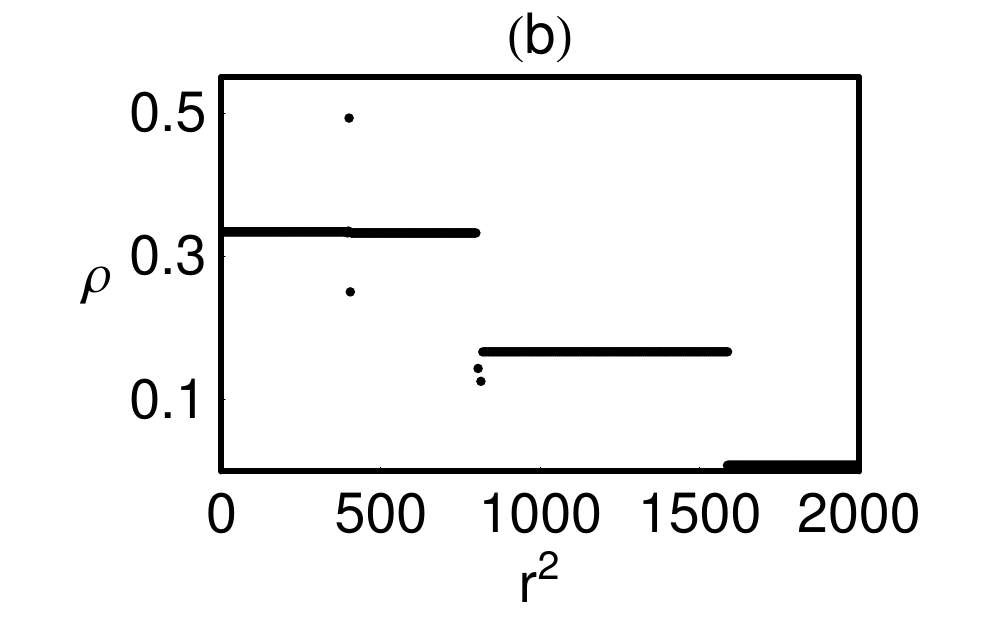}
\includegraphics[width=7cm]{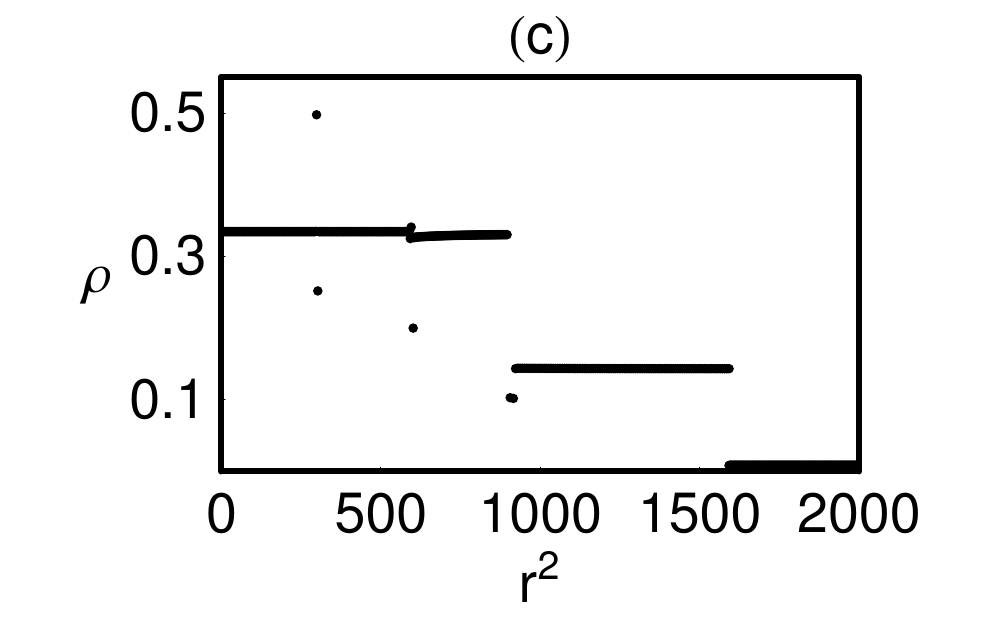}
\end{center}
\caption{Density plots for the matrix ground states
with $1/\nu_{cl}=1/\nu^*+k$, for $k=4$ and $N=400$: 
(a) $\nu^*=2$ (\ref{nu2-sol}); (b) $\nu^*=3$; and (c) $\nu^*=4$.}
\label{fig-drops}
\end{figure}

In Fig.\ref{fig-drops} we plot the densities of the $\nu^*=2$ ground state
(\ref{nu2-sol}), and those of the corresponding $\nu^*=3,4$ states \cite{cr2},
for $N=400$: up to finite-$N$ fluctuations,
they show  two-step uniform densities as anticipated.
We recall that the same droplet shape is found for
the Jain states (\ref{jain-elecstate}), before their
projection to the lowest Landau level \cite{jain}.

In Ref.\cite{cr2} we found a simple argument for the equivalence of the
semiclassical solutions to the matrix wavefunction found in section 5
(cf. (\ref{two-ll})).
We evaluated their polynomial parts $\Phi(\ov{A},\ov{B},\psi)$ 
on the classical solution $\ov{A},\ov{B}$, e.g. (\ref{nu2-sol}),
corresponding to the leading $N\to\infty$  expectation values. 
We then found that the resulting polynomial selfconsistently match 
the single particle occupancies predicted by the classical solution
themselves. These results confirm the validity of the semiclassical
approximation for these matrix ground states.

%-6-----------------------------------------

\section{Electrons from D0 branes in the $g\to\infty$ limit}

\subsection{The matrix theory at $g=\infty$}

In this section we introduce the potential
$V=-(g/2) \Tr [X_1,X_2]^2$ in the Hamiltonian (\ref{mcs-ham})
and perform the $g\to\infty$ limit.
At the classical level, the $V$ potential suppresses the matrix
degrees of freedom different from the eigenvalues, and
projects them out for $g\to\infty$.
Using the Ginibre decomposition of complex matrices
\cite{mehta}:
$X=\ov{U}(\L+R)U$, where $U$ is unitary (the gauge degrees of
freedom), $\L$ diagonal (the eigenvalues) and $R$ complex upper triangular
(the additional d.o.f.),  we find for $N=2$:
\be
V=\frac{g}{8}\Tr\left[X,\ov{X}\right]^2 =
\frac{g}{4}\vert r\vert^4 +
\frac{g}{4}\vert\ov{r}\left(\l_1-\l_2\right)\vert^2\ ,
\qquad
X=\left(\begin{array}{cc} \l_1& r\\ 0 & \l_2 \end{array}\right) .
\label{v-form}
\ee
Thus for large $g$, the variable $r$ is suppressed.
For general $N$, the potential
kills all the $N(N-1)$ degrees of freedom contained in the $R$ matrix.

Let us now discuss the Maxwell-Chern-Simons theory in the
$g=\infty$ limit, i.e. for $R=0$: $X$ and $\ov{X}$ commute among 
themselves and can be
diagonalized by the same unitary transformation,
\be
X=\ov{U}\L U\ ,\quad \ov{X}=\ov{U}\ov{\L} U \ , \qquad
\L = diag\ \left(\l_a\right)\ ,
\qquad
\left[X,  \ov{X}\right] = 0\ .
\label{n-mat}
\ee
The $g=\infty$ theory is analyzed following a different strategy
from that of section 4: we fix gauge invariance, solve the Gauss law
at the classical level and then quantize the remaining variables,
which are the complex eigenvalues $\l_a$ and their conjugate momenta
$p_a$ \cite{poly97}\cite{park}.
We take the diagonal gauge for the coordinates
and decompose the momenta $\Pi,\ov{\Pi}$,
in diagonal and off-diagonal matrices, respectively called $p$  and $\G$:
$ X=\L$ ,$ \P=p+\G$ ,$\ov{\P}=\ov{p}+\ov{\G}$.
The Gauss law constraint (\ref{mcs-gauss}) can be rewritten:
\ba
\left[X,\P \right]+ \left[\ov{X},\ov{\P} \right] &=&
-i\ \B\th + i\ \psi\otimes\ov{\psi}\ ,
\nl
\left(\l_a-\l_b\right) \G_{ab} +
\left(\ov{\l}_a-\ov{\l}_b\right) \ov{\G}_{ab}
&=& -i \left(k\ \d_{ab}- \psi_a\ \ov{\psi}_b\right)\ .
\label{eigen-gauss}
\ea
For $a=b$, this equation implies $\vert \psi_a \vert^2=k$ for any value of $a$:
for $a\neq b$, it completely determines the off-diagonal momenta:
\be
\G_{ab}\ =\ \frac{i k}{2}\
\frac{\ov{\l}_a-\ov{\l}_b}{\vert \l_a-\l_b\vert^2}\ ,
\qquad\quad a\neq b\ .
\label{gamma-val}
\ee
By inserting this back into the Hamiltonian (\ref{mcs-ham}), we find that
diagonal and off-diagonal terms decouple and obtain,
\ba
H_{g=\infty} &=& 2\ \Tr \left[\left(\frac{\ov{X}}{2} -i\ \P\right)
\left(\frac{X}{2} +i\ \ov{\P}\right) \right] \
\nl
 &=& 2\ \sum_{a=1}^N \ \left(\frac{\ov{\l}_a}{2} -ip_a\right)
\left(\frac{\l_a}{2} +i\ov{p}_a\right) \ +\
\frac{k^2}{2}\ \sum_{a \neq b =1}^N\
\frac{1}{\vert \l_a-\l_b\vert^2}\ .
\label{inf-ham}
\ea

Therefore, the theory reduced to the eigenvalues corresponds
to the ordinary Landau problem for N electrons plus an induced
two-dimensional Calogero interaction.
Note also that the matrix measure of integration becomes flat
after incorporating one Vandermonde factor $\D(\l)$ into the wave functions.
The occurrence of the Calogero interaction
is a rather common feature of matrix theories reduced to
eigenvalues: in the present case, the interaction is two-dimensional, owing
to the presence of two Hermitean matrices, and thus it is
rather different from the exactly solvable one-dimensional case
\cite{poly1}\cite{poly5}.

We conclude that the Maxwell-Chern-Simons matrix theory at
$g=\infty$ makes contact with the physical problem
of the fractional quantum Hall effect.
The $e^2/r$ Coulomb repulsion is replaced by the Calogero interaction 
$k^2/r^2$; however, numerical results \cite{laugh}\cite{jain}\cite{numeric}
indicate that quantum Hall incompressible states are rather 
independent of the type of repulsive potential, for large $\B$.
(The specific form of the potential clearly affects the detailed values
of some quantities such as the gap.)

Some remarks are in order:
\begin{itemize}
\item
The physical condition imposed by the Gauss law (\ref{eigen-gauss})
is still that outlined in
section 3.2.1: it forces the electrons to stay apart by
locking their density to the value of the background parameter $k$.
The solution of this constraint is however rather different at
the two points $g=0$ and $g=\infty$:
for $g=0$, it is the geometric, or kinematic, condition of noncommutativity
(\ref{poly-glaw}),
while at $g=\infty$ this is a dynamical condition set by a repulsive
potential with appropriate strength.
\item
Note also that the $g=\infty$ theory is not, by itself, less
difficult than the ab-initio quantum Hall problem: the gap is
non-perturbative and there are no small parameters.
The advantage of embedding the problem into the matrix theory is
that of making contact with the solvable $g=0$ limit, as we discuss in
the next section.
\end{itemize}

%-f6-------------------------------------------
\begin{figure}
\begin{center}
\includegraphics[width=8cm]{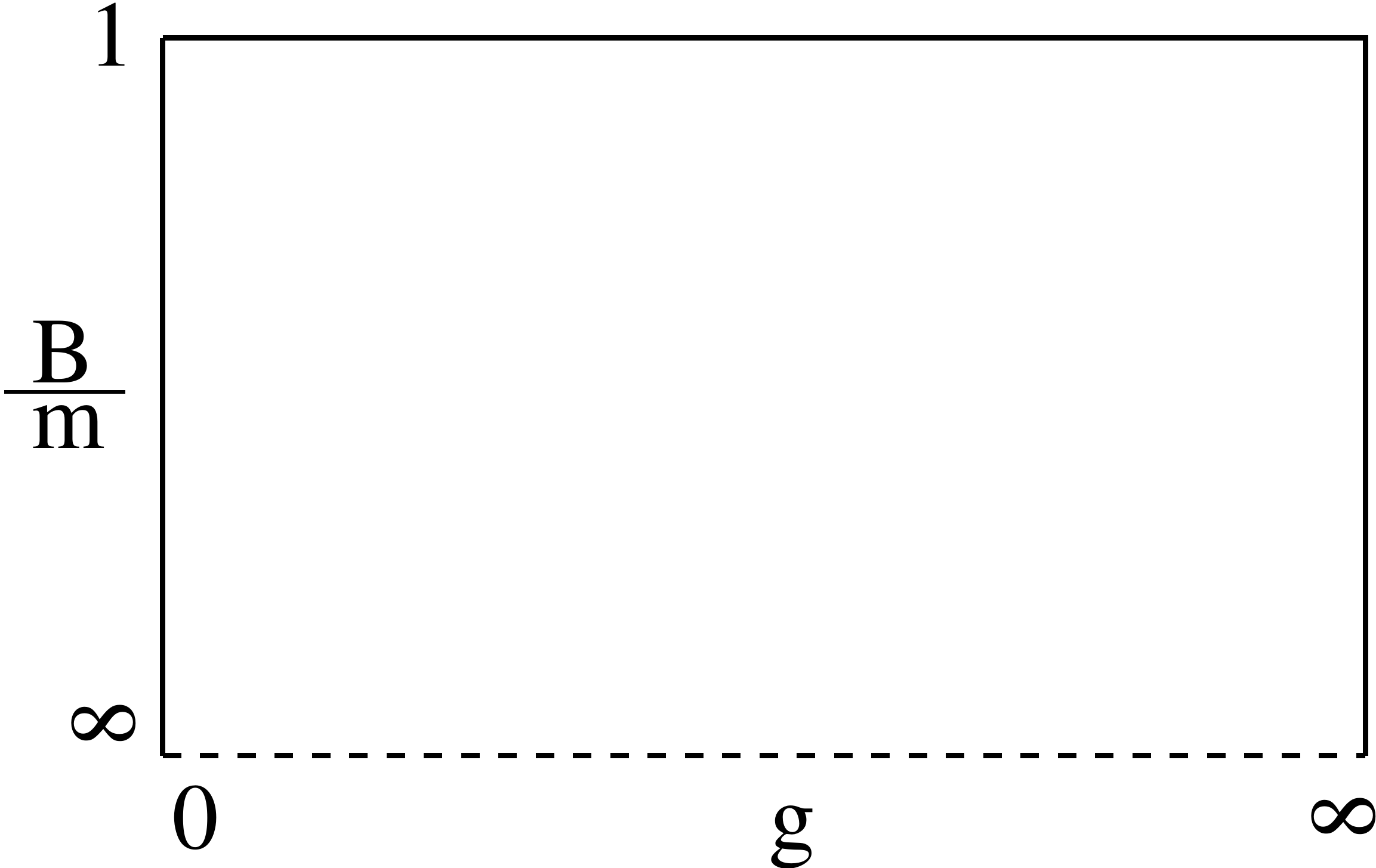}
\end{center}
\caption{Phase diagram of the Maxwell-Chern-Simons matrix theory.
The axes $g=0$ and $g=\infty$ have been discussed in sections
5 and 6, respectively. The Chern-Simons matrix model sits at the
left down corner.}
\label{fig-phase}
\end{figure}

%-6.2------------------------------------------------------

\subsection{Conjecture on the phase diagram}

In Figure (\ref{fig-phase}) we illustrate the phase diagram of the
Maxwell-Chern-Simons matrix theory as a function of its
parameters $\B/m$ and $g$.
The quantized background charge $\B\th=k$ is held fixed
over the diagram together with  average density of the system.

The axes $g=0$ and $g=\infty$ have been discussed in sections
5 and 6.1, respectively. For $g=0$, the theory is
solvable and displays a set of states that are in
one-to-one relation with the Laughlin and Jain ground states
with filling fractions $\nu=m/(mk+1)$.
These non-degenerate states are selected by choosing
the appropriate projection $A^m\ap 0$ and the values of
$k$ and the density (or the angular momentum).
For $g=\infty$, we found that the theory describes a realistic
Hall system, but its ground states are difficult to find.

In ref.\cite{cr1}, we conjectured that the matrix ground states at $g=0$ 
could match one-to-one the phenomenological Jain states that are
good ansatz in the physical limit $g=\infty$ (including the case of
Calogero interaction) \cite{jain}\cite{numeric}: indeed,
the two sets of states become identical in the limit of both $\ov{X},X$ 
diagonal, that is (classically) achieved at $g=\infty$. 
In order to prove this conjecture, 
we would need to consider the evolution of the matrix ground states as the
coupling is varied in between, $0<g<\infty$, and to check that the gap never
vanishes, i.e. that there are no phase transitions in $(B,g)$ plane
 separating the $g=0$ and $g=\infty$ regions
at the specific density values \cite{cr1}.
This conjecture of smooth evolution of matrix Jain states
is indirectly supported by the numerical analyses, showing that the 
Jain wavefunctions are accurate ground states of the $g=\infty$ theory. 
Further support is given by the form of the semiclassical density 
of $g=0$ matrix states that is the qualitatively the same of $g=\infty$
Jain incompressible fluids states.

Let us finally remark that, the limit $\B\to\infty$
cannot be taken at $g=0$, because quasi-particle excitations and
Jain states in the matrix theory have energies of $O(\B)$ and
would be projected out.  Instead, the limit $\B=\infty$ can surely
be taken in the $g=\infty$ physical theory
(holding $k=\B\th$ fixed), because the fractional
quantum Hall states are known to remain stable.
This implies that the two limits are ordered: the correct sequence is
$\lim_{\B\to\infty}\ \lim_{g\to\infty}\Psi$, and the opposite choice
is cut out in the phase diagram of Fig.\ref{fig-phase}.

%-7-----------------------------------
\section{Conclusions}

We have reviewed the description of the fractional quantum Hall effect
given by gauge matrix theories, that provide one realization
of the composite-fermion correspondence.
In particular, the Maxwell-Chern-Simons theory, supplemented by
certain projections of states, reproduces
the Jain hierarchical construction of ground state wavefunctions.
These results support the idea that the fractional Hall states
should be uniquely characterized by algebraic 
conditions and gauge invariance, rather than by detailed dynamics, because
they are exceptionally robust and universal.

The study of the phase diagram of the matrix theory is clearly necessary
to make better contact between the nice results ($g=0$)  and the
physical regime ($g=\infty$), upon varying the potential
$V=g\ \Tr[\ov{X},X]^2\ $. We plan to study the evolution 
of matrix ground states for $g>0$ by including 
the quartic potential in the semiclassical analysis within the
mean-field approximation.

One point to develop is the study of edge excitations of matrix Jain states
\cite{rod} and the comparison with the conformal field theory descriptions
\cite{jainedge}: in particular, the realization of the
$SU(n)$ symmetry, for $\nu=n/(2kn+ 1)$, that is still debated
\cite{cz}.
Another open problem is the derivation of the fractional statistics
of quasiparticles in the matrix theory setting.
Both issues require an improvement of the $A^n\ap 0$ projection 
that could better handle excitations above the ground state.

\ack

The authors would like to thank the hospitality of the G. Galilei 
Institute for Theoretical Physics, Florence.
This work was partially funded by the ESF programme 
{\it INSTANS: Interdisciplinary Statistical and Field Theory Approaches 
to Nanophysics and Low Dimensional Systems}, and by the MUR grant
{\it Fisica Statistica dei Sistemi Fortemente Correlati all'Equilibrio 
e Fuori dall'Equilibrio}.

%-r---------------------------------

\end{document}